\documentclass[a4paper,twocolumn,11pt]{quantumarticle}
\pdfoutput=1

\usepackage[colorlinks, citecolor=blue]{hyperref}
\usepackage[numbers,sort&compress]{natbib}
\usepackage{graphicx}
\usepackage{amsmath}
\usepackage{amssymb}
\usepackage[english]{babel}
\usepackage{color}
\usepackage[version=4]{mhchem}
\usepackage{dsfont}
\usepackage{multirow}
\usepackage{booktabs}
\usepackage{array}
\usepackage{stmaryrd}
\usepackage{physics}
\usepackage{bm}
\usepackage{comment}
\usepackage{braket}
\usepackage[normalem]{ulem} 

\usepackage{soul}

\usepackage{textcomp}

\usepackage{float}
\makeatletter
\let\newfloat\newfloat@ltx
\makeatother

\usepackage{algorithm}
\usepackage{algpseudocode}

\graphicspath{{./Figures/}}

\newcommand{\UIUC}{
    Department of Physics,
    The University of Illinois at Urbana-Champaign,
    Urbana, IL 61801, USA
}

\newcommand{\UW}{
    Department of Physics,
    University of Wisconsin -- Madison,
    Madison, WI 53706, USA
}

\newcommand{\red}[1]{{\color[rgb]{0,0,0}{#1}}}

\renewcommand{\cite}[1]{\mbox{\citep{#1}}}

\addto\captionsenglish{}
\addto\captionsenglish{\renewcommand\thefigure{\arabic{figure}}}

\newcommand{\Output}{\item[\textbf{Output:}]}

\begin{document}

\title{Efficient algorithms for quantum chemistry on modular quantum processors}
\author{Tian Xue}\email{tianxue2@illinois.edu}
\affiliation{\UIUC}
\author{Jacob P. Covey}\email{jcovey@illinois.edu}
\affiliation{\UIUC}
\author{Matthew Otten}\email{mjotten@wisc.edu}
\affiliation{\UW}

\begin{abstract}
Quantum chemistry is a promising application of future quantum computers, but the requirements on qubit count and other resources suggest that modular computing architectures will be required. We introduce an implementation of a quantum chemistry algorithm that is distributed across several computational modules: the distributed unitary selective coupled cluster \red{algorithm} (dUSCC). We design a packing scheme using the pseudo-commutativity of Trotterization to maximize the parallelism while optimizing the scheduling of all inter-module gates around the buffering of inter-module Bell pairs. We demonstrate dUSCC on a 3-cluster $(\text{H}_4)_3$ chain and show that it naturally utilizes the molecule's structure to reduce inter-module latency. We show that the run time of dUSCC is unchanged with inter-module latency up to \red{$\sim35\times$} slower than intra-module gates in the $(\text{H}_4)_3$ while maintaining chemical accuracy. dUSCC should be ``free'' in the weakly entangled systems, and the existence of ``free'' dUSCC can be found efficiently using classical algorithms. This new compilation scheme both leverages pseudo-commutativity and considers inter-module gate scheduling, and potentially provides an efficient distributed compilation of other Trotterized algorithms.
\end{abstract}
\maketitle

\section{Introduction}
Quantum resource estimates for utility-scale problems, including those in chemistry, materials, and factoring estimate that $\sim$1-10 million quantum bits (qubits) are expected to be required for quantum computers to reach their full potential~\cite{Gidney2021,Beverland2022,otten2024quantum,watts2024fullerene,nguyen2024quantum,bellonzi2024feasibility,Gidney2025,Zhou2025}. Although only several thousand ``perfect" logical qubits are required, the propensity of errors in quantum systems requires the use of quantum error correction (QEC) which incurs large overhead resource costs. The leading quantum computing hardware modalities have scaling roadmaps that will carry us from thousands of physical qubits today to $\sim$$10^4$-$10^5$ qubits via brute-force scaling of existing methodologies. However, for many platforms, approaching and surpassing the million-qubit level will require a paradigm shift.

\begin{figure}[t!]
	\includegraphics[width=\linewidth]{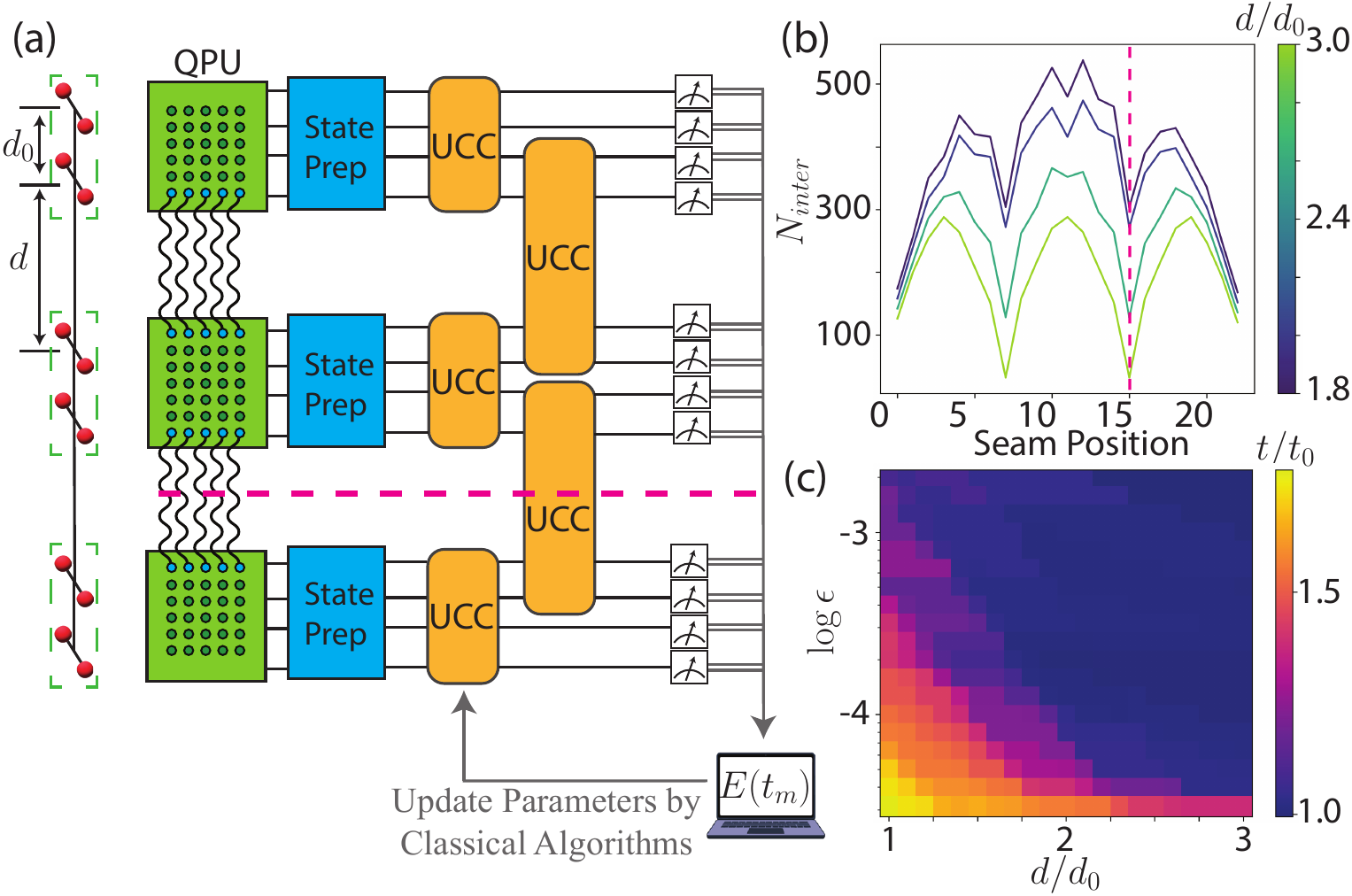}
	\caption{
        \textbf{Overview of the modular architecture}. (a) The dUSCC circuit of the $(\text{H}_4)_3$ chain with inter-cluster separations three times the intra-cluster separations: $d = 3d_0$. The dUSCC circuit is distributed to three interconnected QPUs. Classical ansatzes are prepared on each QPU followed by the VQE \red{with inter-module dUSCC ansatz}. (b) The number of inter-module CNOTs ($N_{inter}$) of dUSCC of the $(\text{H}_4)_3$ chain with $\epsilon = 10^{-3}$ at different positions of the seam between modules. $N_{inter}$ reaches its minimum where the entanglement between H$_2$ is weakest. (c) The circuit time ratio of the delayed dUSCC ($t$) and undelayed dUSCC ($t_0$) at different cluster separations $d/d_0$ and different selection accuracies $\epsilon$. The dUSCC circuit is delayed due to the buffering of Bell pairs. There is a clear boundary between the ``strongly" entangled phase (yellow) and the ``weakly" entangled phase (purple) depending on the entanglement between modules, and the best trade-off between accuracy and circuit time is at the phase boundary.} 
    \phantomsection
    \label{fig:schematic}
\end{figure}

This transformation is expected to be accompanied by the need for modularity, where large-scale quantum processors are composed of many inter-connected modules across which the quantum algorithm is distributed \cite{Wehner2018, Niu2023, Wu2025, Ramette2024, Vazquez2024}. It is generally assumed that the quantum interconnects between the modules will not perform as well in terms of rate and/or fidelity as intra-module operations. Hence, as the community strives to build such modular interconnects, a natural question is: what performance level is required? The answer to this question largely depends on the target application and the flexibility of compiling and scheduling the specific algorithm across the modular hardware. For example, it is well known that certain quantum adder sub-routines for factoring algorithms can be efficiently cut into modules with relatively few inter-module operations required~\cite{Gidney2019,Monroe2014,Pattison2024}. However, the other sub-routines in, e.g., Shor's algorithm such as quantum phase estimation are not known to be efficiently distributable.

Here, we focus on the use of modular quantum processors for quantum chemistry -- one of the most promising long-term applications of quantum computing~\cite{Cao2019,McArdle2020}. Specifically, we focus on the task of finding the ground state (up to chemical accuracy) of a large molecule \red{using unitary selective coupled cluster~\cite{Otten2022}, an early-fault tolerant quantum algorithm combining QPE and VQE. Other quantum algorithms with provable convergence (such as qubitization) are very promising for fully fault-tolerant quantum computers~\cite{Babbush2018, Low2025, Low2019}, but the flexibility and resource reduction of USCC, building upon VQE-like algorithms, allow it to be a promising path for quantum simulation in the near future}. We assume molecules are in the Born-Oppenheimer approximation where nuclear degrees of freedom are frozen. We show that modular architectures are uniquely well suited for this application because 1) many large molecules are naturally grouped into clusters with relatively weak but non-trivial connectivity between them, and 2) fermionic Hamiltonians offer flexible scheduling due to the pseudo-commutativity of the Trotterization. 

We focus on the $(\text{H}_4)_3$ chain, where each of the three H$_4$ systems is represented with a quantum processing unit (QPU) (see Fig.~\ref{fig:schematic}). We demonstrate that, to achieve 
the standard benchmark of chemical accuracy (1.6 mHa), our algorithm can tolerate inter-module gates up to \red{$\sim35\times$} slower than intra-module gates without increasing the runtime. Our resource scaling estimates suggest that the savings of our algorithm are roughly linear in the number of modules compared with a naive compilation, which becomes substantial for large molecular systems. This work adds quantum chemistry to the list of applications for modular quantum computers and illustrates the benefits of the pseudo-commutativity of most Trotterized algorithms for efficient quantum algorithm compilation.

\begin{figure*}[t]
	\includegraphics[width=\linewidth]{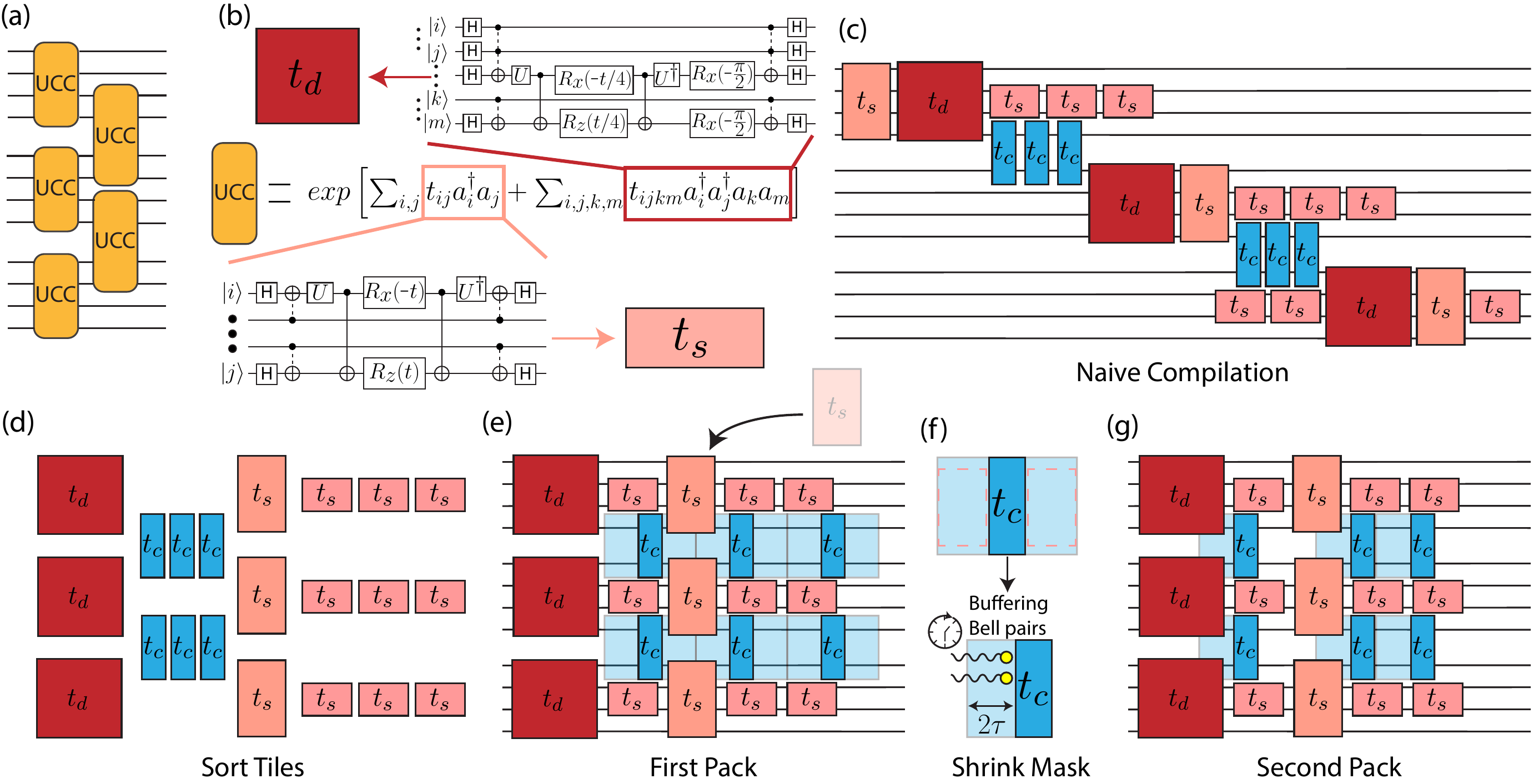}
	\caption{
        \textbf{Optimization of the modular circuit}. \red{(a) The dUSCC ansatz can be decomposed to multiple inter-module ansatzes (b) dUSCC ansatzes are loaded to the circuit by JW transformation, Trotterization and simplified by ZX calculus. All terms in the dUSCC are compiled to pseudo-commutative circuit tiles. For illustration, single hopping tiles (tiles indexed by $t_s = t_{ij}$) are compiled to light red tiles, double hopping tiles (tiles indexed by $t_d = t_{ijkm}$) are compiled to dark red tiles, and the controlled hopping tiles (tiles indexed by $t_c = t_{inm}$) are compiled to blue tiles. Controlled hoppings are special double hoppings with $j=k\equiv n$. The exact circuits with ZX simplification are shown in Fig. \ref{fig:zx proof}. (c) The naive compilation of the USCC ansatz without consideration of inter-module latency and geometry of molecules (d) In the dUSCC, all circuit tiles are first sorted by their heights. (e) Using the pseudo-commutativity of circuit tiles, the circuit is packed to maximize the intra-module parallelism. All inter-module tiles are masked with additional width to simulate the inter-module latency. Intra-module tiles can fit into the mask due to the parallelism between the buffering of inter-module Bell pairs and the intra-module operations. The mask is set as $\tau_m \geq \tau$ to ensure there exists a tile that fits the mask. (f) The mask of inter-module tiles is shrunk to $2\tau$ to simulate the exact buffering time required for each tile. (g) Tiles are packed again with the shrunk mask to remove the additional cost from maximizing the parallelism between inter- and intra-module tiles from the first pack. In the systems with sparse inter-module entanglement, all inter-module communications are parallel to intra-module operations.}}
    \phantomsection
    \label{fig:packing}
\end{figure*}

\section{Theory and Methods}
In this study, we focus on the unitary selective coupled cluster (USCC) algorithm, which has been demonstrated as an efficient method for generating compact ansatzes that reduce the gate depth while maintaining accuracy for quantum chemistry~\cite{fedorov2022unitary,mitra2024localized,Otten2022}. The algorithm has three main components. First, the localized classical solutions are loaded onto QPUs through direct initialization (DI) or quantum phase estimation (QPE)~\cite{DCunha2024}. \red{Second, the 
 USCC ansatzes are compiled to qubit circuits by the Jordan-Wigner (JW) transformation and Trotterization~\cite{Berry2006, Ovrum2007}.}  Lastly, the ground state energy is approximated by a variational quantum eigensolver (VQE)~\cite{McClean2016,Kandala2017}, and the variational parameters are updated accordingly [See Fig. \ref{fig:schematic}(a)]. A \red{naive} compilation of USCC using JW transformation and Trotterization is illustrated in Appendix \ref{USCC loading}. \red{A more compact compilation using ZX calculus \cite{Cowtan2020} is in Appendix \ref{appendix: zx proof}}

As noted in the introduction, simulations of large molecules may require a distributed implementation of USCC (dUSCC) across many computational modules. USCC only selects important hopping terms with energy gradients $\frac{\partial E}{\partial t_\beta}|_{t_\beta = 0}\geq \epsilon$ to reduce the complexity of the VQE. Since the dominant interactions between electrons are Coulomb interactions, the energy gradient decreases as the hopping distance increases.

In dUSCC, the entanglement between modules determine the required inter-module communication and latency. In the $(H_4)_3$ chain, the number of inter-module CNOTs is minimal since \red{the entanglement is weaker at the stretched bonds}.
\red{The entanglement between clusters also decreases at larger $\epsilon$}.  The long-range interactions are filtered \red{at higher $\epsilon$} due to their low contribution to the energy. Therefore, the required number of inter-module CNOT gates drops rapidly as the distance between clusters increases and the entanglement between clusters becomes weaker as shown in Fig. \ref{fig:schematic}(c).
  
The weak entanglement between modules still requires inter-module communications and may significantly delay the circuit. We assume that all qubits are ``perfect", implying that we are operating at the logical level with an error-correcting code of distance of $d$, but we ignore time for error correction. We also assume that all QPUs are connected by Bell pairs between physical qubits, and that these Bell pairs are consumed when performing inter-module gates. To note, most hardware platforms currently generate these remote Bell pairs serially. Hence, while intra-module transversal CNOT gates at the logical level can be performed at nearly the same rate as physical CNOT gates (at least with transportable atom array hardware~\cite{Bluvstein2022,Bluvstein2023}), inter-module operations at the logical level are much less efficient. Specifically, an inter-module logical CNOT gate performed via lattice surgery requires $d$ shared Bell pairs while a transversal CNOT gates requires $d^2$ shared Bell pairs~\cite{Horsman2012, Erhard2021, Wan2019}. This overhead incurs a substantial latency if the Bell pairs are generated serially.
For added context, a detailed implementation of intra/inter-module operations on the neutral atom array platform using lattice surgery is described in Appendix \ref{benchmark}.

Our compilation of the dUSCC algorithm is inspired by classical high-performance computing (HPC): we maximize parallelism among modules while hiding the inter-module communication behind intra-module calculations~\cite{Dillon2010, OpenMP1998,bianco2013, otten2016mpi,raffenettiMPI2017}. We mainly focus on two types of parallelism: 1) parallelism between modules, and 2) parallelism between intra-module operations and buffering of Bell pairs. The dUSCC circuit is dominated by CNOT gates with sparse single-qubit gates. \red{We assume the circuit time only depends on these two types of gates and remove other Clifford gates and Pauli gates in our gate/depth count.} In the rest of the paper, we will use this effective circuit depth to represent the circuit time. The single-qubit gates on the logical qubit level can be applied by the magic state injection, partially fault-tolerant gates \red{or direct gate decomposition}~\cite{Akahoshi2024, Ito2025, Ross2016}, and this can happen in parallel with the buffering of Bell pairs. \red{In this paper, we assume the logical single qubit rotation at arbitrary angle is decomposed to $\text{Clifford}+T$ gates, and each single qubit rotation is $10\times$ more expensive than a logical CNOT. More analysis on the impact of cost of single qubit rotations to dUSCC is in Appendix \ref{Appendix: Double Packing}}. The additional time required for single-qubit operations (such as magic-state cultivation, distillation and injection, quantum error correction, etc.) provides more opportunities to buffer Bell pairs and further enhances the parallelism in dUSCC. 

Although USCC circuits can be compiled naively by Qiskit~\cite{Barkoutsos2018}, this compilation ignores the parallelism in the distributed architecture and the associated hardware limitations. dUSCC is compiled to parallelize tasks using the pseudo-commutativity of Trotterization.
Since the Trotterization error comes from the commutator, the order of magnitude of the Trotter error is unchanged under reordering of each $\exp\left(it_kH_k\right)$~\cite{Childs2021}. \red {This pseudo-commutativity is explained in more detail in Appendix \ref{appendix: pseudo-commutativity}}. 

\red{The USCC ansatz can be decomposed to multiple inter-module ansatzes [Fig. \ref{fig:packing}(a)], and each inter-module ansatz can be compiled to circuit tiles using Trotterization and the JW transformation, and then simplified by ZX calculus~\cite{Cowtan2020} [Fig. \ref{fig:packing}(b)]. Tiles can be naively compiled to circuits (such as Qiskit) without consideration of geometry of molecules and inter-module architecture [Fig. \ref{fig:packing}(c)].  Using the pseudo-commutativity of circuit tiles, we can maximize the parallelism of the dUSCC between modules by solving a classical 1+1D tile packing problem. In dUSCC, all circuit tiles are sorted by their heights and packed using the greedy algorithm [Fig. \ref{fig:packing}(d, e)]. Inter-module tiles are masked with additional width to simulate the inter-module latency, where only the intra-module tiles can fit into the mask. The mask is first set to $\tau_m \geq \tau$ to ensure there exists an intra-module tile that fits the mask [Fig. \ref{fig:packing}(e)]. The inter-module mask is then shrunk to $\tau$ and tiles are packed again to remove the additional cost to maximize the inter/intra parallelism [Fig. \ref{fig:packing}(f)]. In sufficiently weakly entangled systems, communications can be perfectly parallel to calculations as shown in Fig. \ref{fig:packing}(g).  More details about the heuristic greedy double packing scheme are in Appendix \ref{Appendix: Double Packing}.}

\begin{figure}[t!]
	\includegraphics[width=\linewidth]{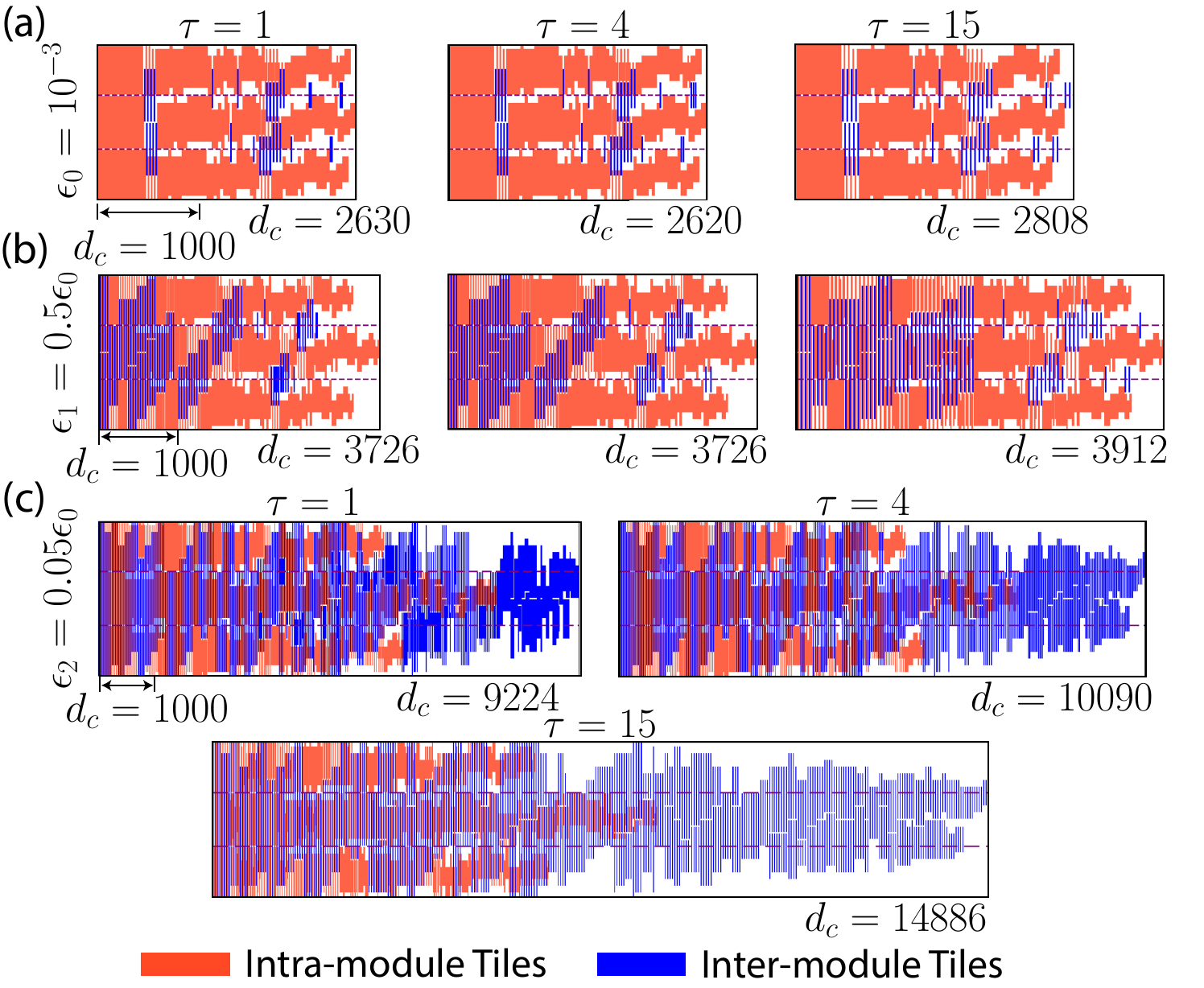}
	\caption{\textbf{The abstract circuit of the 3-module $(\text{H}_4)_3$ chain with $d = 3d_0$ at different costs of inter-module gates}. The effective circuit depth \red{($d_c$)} represents the circuit time. The width of the figures also represents the effective circuit depth with calibrations at the left bottom corner. The effective circuit depth only counts the depth of CNOTs \red{and single qubit rotations at arbitrary phases (assumed to be 10$\times$ expensive as a CNOT)}.
        (a) The dUSCC circuit packing of the $(\text{H}_4)_3$ at $\epsilon_0 = 10^{-3}$. The dUSCC circuit is first compiled to tiles and packed by the double packing algorithm to maximize the parallelism between Bell pair buffering and intra-module operations. At $\tau = 4,15$, inter-module tiles are separated to buffer Bell pairs for the next inter-module tiles. 
        (b) The selection criterion is then lowered to $0.5\epsilon_0$. The inter-module entanglement becomes stronger, but the inter-module communications can still be buffered behind intra-module operations. At $\tau = 15$, inter-module tiles are evenly distributed over space to optimize the parallelism. The modularization of dUSCC is ``free'' in all plots in (a) and (b).
        (c) The selection criterion is lowered to $0.05\epsilon_0$ and the $(\text{H}_4)_3$ chain becomes extensively and strongly entangled. The number of inter-module tiles exceeds all available parallel space of intra-module tiles, so there is a inter-module gate tail at the end of the circuit. Increasing $\tau$ will linearly expand the tail, and the circuit can no longer be optimized by any packing scheme.
    }
    \phantomsection
    \label{fig:packing plot}
\end{figure}

\section{Results}
In Fig. \ref{fig:schematic}(a), we consider a simplistic model of six weakly interacting H$_{2}$ molecules with various distances between them (green box). \red{dUSCC applied to more complicated molecules is in Appendix \ref{appendix:more molecules}.}  We assume the $(\text{H}_2)_6$ chain are clustered into three groups that each contains two H$_2$ molecules, i.e. $(\text{H}_4)_3$, where $d_0$ is the distance between the two intra-cluster H$_2$'s and $d$ is the distance between adjacent inter-cluster H$_2$'s. The state preparations are localized in each cluster. \red{The active space for each electron includes two orbitals, and the dUSCC of $(\text{H}_4)_3$ contains 24 qubits}. The simulation uses the minimal STO-3G (Slater-type orbitals) basis to ensure the largest computationally feasible H-chain~\cite{pople1965}. The \red{$\text{H}_4$-chain} with 3 clusters is the smallest chain to efficiently simulate a translation invariant H$_2$-chain, and we expect the result only changes slightly in the infinite H$_2$-chain as shown in Appendix \ref{extended H-chain}. 

In Fig. \ref{fig:packing plot}, we simulate the dUSCC packing of the $(\text{H}_4)_3$ chain with $d = 3d_0$ at different selection criteria $\epsilon$ and costs of inter-module Bell pairs: $\tau \equiv \frac{t_{\text{Bell pairs}}}{t_{intra-CNOT}}$.
Red tiles are intra-module circuit tiles and blues are inter-module tiles. The effective circuit depth \red{($d_c$)} represents the dUSCC circuit time as noted in the theory, \red{and we assume the single rotation gate at arbitrary phases is $10\times$ as expensive as a logical CNOT.} In Fig. \ref{fig:packing plot}(a), we start at $\epsilon_0 = 10^{-3}$, which approximates the ground state of the H$_2$-chain \red{within} the chemical accuracy~\cite{Verma2025}. The circuit is packed using the double packing algorithm to maximize the parallelism between inter/intra-tiles.
Inter-module tiles are masked to include the inter-module latency. Buffering of Bell pairs is perfectly parallel to intra-module calculations, and the circuit time remains unchanged. The selection criterion is then lowered to $\epsilon_1 = 0.5\epsilon_0$ to include more hoppings in Fig. \ref{fig:packing plot}(b). The circuit has more entanglement, but the circuit time still remains unchanged as $\tau$ increases. When $\tau = 15$, all inter-module tiles are evenly distributed throughout the circuit, and inter-module communications are maximally parallel with intra-module calculations. However, for $\epsilon_2 = 0.05\epsilon_0$, the circuit generates stronger entanglement as shown in Fig. \ref{fig:packing plot}(c).
The large inter-module tail suggests that the Bell pair generation latency over-saturates all parallelism from intra-module operations. As $\tau$ increases, this tail expands linearly with the inter-module latency, leading to a linear increase of the circuit time.

\begin{figure}[H]
    \includegraphics[width=\linewidth]{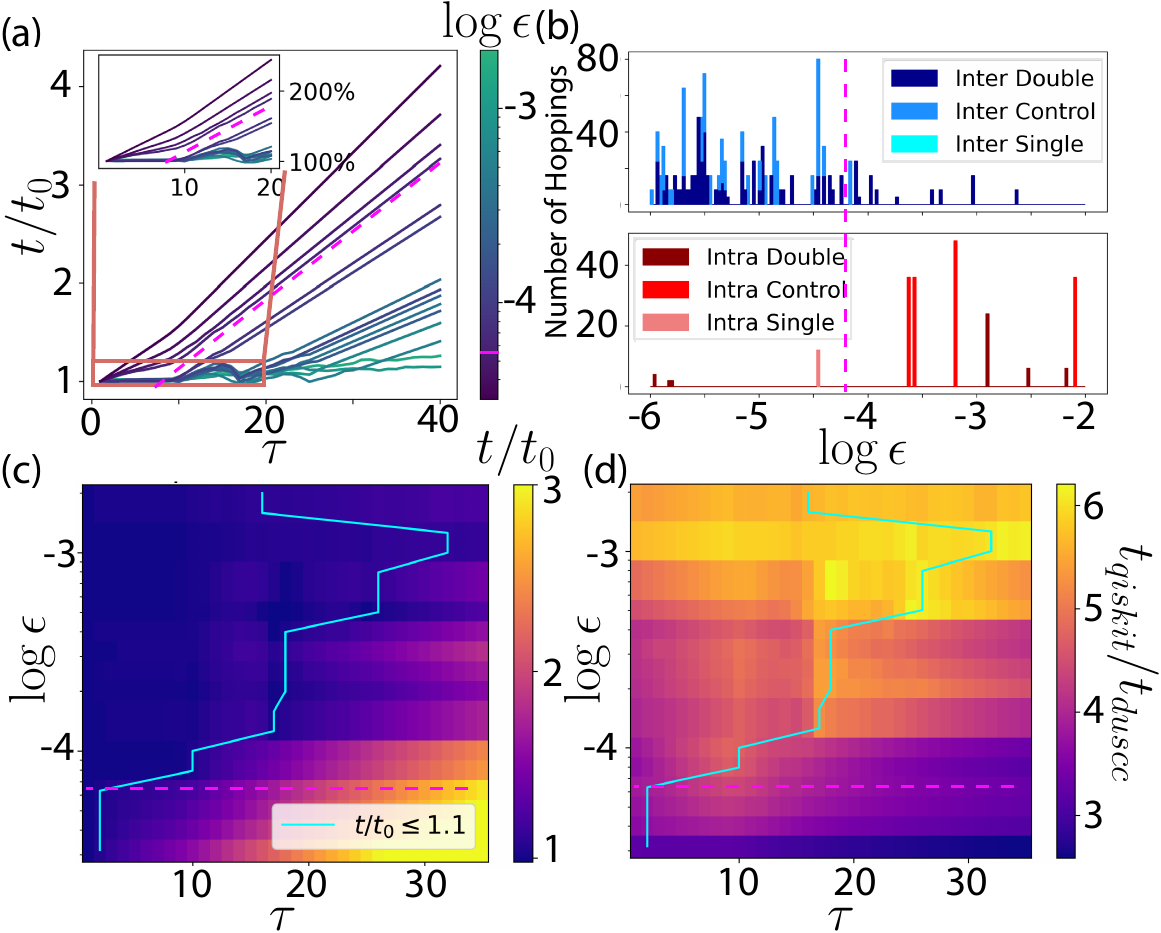}
	\caption{
        \textbf{Results}. \red{All results are from the dUSCC simulation in the $(H_4)_3$ at $d = 3d_0$} (a) dUSCC circuit time compared to the no-latency case, $t/t_0$, with different inter-module gate costs $\tau$ and different selection criteria $\epsilon$. The dUSCC circuit time increases linearly with the inter-module gate costs. The slope depends on the selection criteria $\epsilon$, and the modularization of USCC circuit is ``free'' below the threshold $\tau$ (red box) in weakly entangled systems. The dashed pink line suggests the boundary between the existence of ``free'' modularization dUSCC. The ``free'' modularization only exists on the right of the dashed line, while the modularization always delays the circuit on the left of the dashed line.
        (b) The hopping diagram of $(\text{H}_4)_3$ over the selection criteria. There also exists a boundary between inter-module hoppings and intra-module hoppings, and this boundary captures the ``free'' modularization threshold in (a). One can calculate the hopping diagram in $O(\text{poly}(N_{orbital}))$ time, so the classical algorithm can efficiently find the existence of ``free'' modularization at a selection $\epsilon$. (c) The phase diagram of the delayed dUSCC time. The blue line captures the transition from the ``free'' dUSCC with $t/t_0 < 1.1$ to the delayed dUSCC. The existence of the ``free'' modularization is also consistently captured by the pink line $\epsilon = 10^{-4.2}$. (d) The time ratio between the dUSCC compilation and the Qiskit compilation at. Our compilation shows its best advantages at the blue boundary in (c).
    }
    \phantomsection
    \label{fig:result}
\end{figure}

We time the dUSCC of $(\text{H}_4)_3$ at $d=3d_0$ with more inter-module latencies and choices of the selection criteria as shown in Fig. \ref{fig:result}. Fig. \ref{fig:result}(a) verifies the linear relation between the inter-module latency and dUSCC circuit time. The slope increases as the entanglement between $\text{H}_4$ is stronger. The red box in Fig. \ref{fig:result}(a) indicates that the modularization of the dUSCC circuit does not delay the algorithm if $\tau$ is below the threshold. The pink dashed line labels the boundary of the existence of a threshold, and it vanishes at $\epsilon \leq 10^{-4.2}$ as inter-module entanglement becomes dominant. The boundary of the existence of the threshold $\epsilon = 10^{-4.2}$ can approximate the ground state energy of H$_2$-chain below $\sim 1/100$ chemical accuracy, and $\epsilon \leq 0.001$ estimates ground state energy of the H-chain below the chemical accuracy~\cite{Verma2025}, which has threshold $\tau_t \geq 35$ .

Fig. \ref{fig:result}(b) illustrates the hopping diagram around different $\epsilon$. 
There is a clear boundary between the intra- and inter-module hoppings captured by the same pink dashed line $\epsilon = 10^{-4.2}$, and the threshold only exists when the number of inter-module entanglement is sparse such that intra-module operations can accommodate all buffering of Bell pairs. This hopping diagram can be computed in polynomial time classically, and one can quickly access the amenability of a molecular system to modularization at relatively low cost~\cite{Verma2025}.

Fig. \ref{fig:result}(c) is the phase diagram of the dUSCC circuit time. The blue curve labels the ``free'' modularization transition where $t/t_0 \leq= 1.1$. The $10\%$ relaxation is because our packing scheme is not optimal, and the circuit time has small fluctuations even when the inter-module latency is below the threshold. The ``free'' modularization transition does not decrease monotonically as $\epsilon$ decreases since sometimes lowering $\epsilon$ only introduces more intra-module hoppings, which gives more space to parallelize inter-module tiles and thus lifts transition points. The boundary of the existence of ``free'' modularization is also captured by the pink dashed line $\epsilon=10^{-4.2}$. 

We also time the dUSCC circuit of $(\text{H}_4)_3$ at different distances at $\tau = 4$ as shown in Fig.~\ref{fig:schematic}(c). As the separations between clusters increase, the dUSCC time decreases significantly with a clear region of ``free'' modularization.

Lastly, we compare the dUSCC compilation with the Qiskit compilation in Fig. \ref{fig:result}(d). In the $(\text{H}_4)_3$ chain, the Qiskit circuit is at most $3\times$ longer than the dUSCC circuit, and this ratio increases linearly with $N_{modules}$ as shown in Appendix \ref{extended H-chain}. As demonstrated in Appendix~\ref{app:FeMoco}, utility-scale problems could require $O(50)$ modules. Our compilation has the most advantage over Qiskit near the transition points  shown by the blue curve in Fig. \ref{fig:result}(c). 

\section{Concluding discussion}
We introduced the dUSCC compilation that optimizes both intra-module calculations and inter-module communications. The parallelism between modules is optimized using circuit tile packing by leveraging the pseudo-commutativity of Trotterization. The parallelism between intra-module operations and buffering of Bell pairs is optimized by separating inter-module tiles. This simultaneous optimization significantly accelerates the circuit time compared with the naive compilation. The dUSCC circuit is at most $\sim6\times$ faster than the Qiskit circuit in $(\text{H}_4)_3$ \red{where $\sim 2\times$ saving is from the ZX compilation, and $N_{cluster}=3$ saving by utilizing the translation invariant geometry of $(H_4)_3$}. This acceleration grows linearly with the number of modules \red{as shown in Appendix \ref{extended H-chain}}. 

The dUSCC circuit manifests the entanglement structure of molecules, which enables the dUSCC to be an efficient algorithm for weakly connected molecules. The modularization of dUSCC is ``free'' below the threshold of at most \red{$\tau_t =35$} in $(\text{H}_4)_3$, and the circuit time grows linearly with the cost of inter-module communications at $\tau$ above the threshold. However, dUSCC cannot provide an exact solution to complicated molecules. At strict selection criteria $\epsilon$, modular systems become strongly and extensively entangled, and the inter-module communications over-saturate the intra-module calculations. In the strongly entangled systems, dUSCC circuit is still significantly faster than the Qiskit circuit, and this advantage grows linearly with the number of modules, but the distribution of dUSCC is not ``free''. Nevertheless, dUSCC can still efficiently approximate the ground state energy within chemical accuracy, and the existence of ``free'' modularization in complicated molecules can be found in polynomial time using classical algorithms.

Our circuit tile compilation is not limited to molecular systems and the dUSCC algorithm. For other Trotterized algorithms, their parallelism can also be enhanced by its pseudo-commutativity using the circuit tile compilation. \red{In this paper, we focused on the hydrogen chain, as it has been previously used as a benchmark systems to compare many methods~\cite{PhysRevX.7.031059} and can be naturally formulated as coupled fragments. Results in Appendix~\ref{appendix:more molecules} demonstrate our method's ability to generalize to more complicated molecules and still provide significant reduction in latency costs on modular architectures. 
Various classical methods, including DMRG~\cite{sawaya2022constructing} and others~\cite{PhysRevX.7.031059}, have been successfully applied to simulate hydrogen chains. However, on more complicated molecules, 
quantum algorithms have been estimated to provide run-time advantage over DMRG~\cite{bellonzi2024feasibility}, and we expect dUSCC to be 
even faster on modular systems than quantum algorithms previously studied.}
We also expect that dUSCC can efficiently simulate other 1D lattice many-body systems with localized periodicity such as the Su–Schrieffer–Heeger (SSH) model~\cite{SSH1979}. Different from DMRG requiring short-range entanglement, dUSCC only requires sparse inter-module hoppings compared with intra-module hoppings but still allows complicated intra-module interactions, which is exponentially prohibitive in DMRG~\cite{Catarina2023}.  As more efficient inter-module connections and larger fault-tolerant quantum computers are developed, we expect our compilation can provide an efficient implementation of complicated many-body simulation of large molecules and materials.\\

\section{Acknowledgments}
We acknowledge the Covey Lab for stimulating discussions. J.P.C. and M.O. acknowledge funding from the NSF QLCI for Hybrid Quantum Architectures and Networks (NSF award 2016136). J.P.C. also acknowledges support from NSF award 137642 and the AFOSR Young Investigator Program (AFOSR award FA9550-23-1-0059). 

\bibliographystyle{plainnat}
\bibliography{library}

@article{Kandala2017,
author = {Kandala, Abhinav and Mezzacapo, Antonio and Temme, Kristan and Takita, Maika and Brink, Markus and Chow, Jerry M. and Gambetta, Jay M.},
doi = {10.1038/nature23879},
issn = {0028-0836},
journal = {Nature},
month = {sep},
number = {7671},
pages = {242--246},
title = {{Hardware-efficient variational quantum eigensolver for small molecules and quantum magnets}},
url = {https://www.nature.com/articles/nature23879},
volume = {549},
year = {2017}
}

@article{McClean2016,
author = {McClean, Jarrod R and Romero, Jonathan and Babbush, Ryan and Aspuru-Guzik, Al{\'{a}}n},
doi = {10.1088/1367-2630/18/2/023023},
issn = {1367-2630},
journal = {New J. Phys.},
month = {feb},
number = {2},
pages = {023023},
title = {{The theory of variational hybrid quantum-classical algorithms}},
url = {https://iopscience.iop.org/article/10.1088/1367-2630/18/2/023023},
volume = {18},
year = {2016}
}

@article{McArdle2020,
author = {McArdle, Sam and Endo, Suguru and Aspuru-Guzik, Al{\'{a}}n and Benjamin, Simon C. and Yuan, Xiao},
doi = {10.1103/RevModPhys.92.015003},
issn = {0034-6861},
journal = {Rev. Mod. Phys.},
month = {mar},
number = {1},
pages = {015003},
title = {{Quantum computational chemistry}},
url = {https://link.aps.org/doi/10.1103/RevModPhys.92.015003},
volume = {92},
year = {2020}
}

@article{Cao2019,
author = {Cao, Yudong and Romero, Jonathan and Olson, Jonathan P. and Degroote, Matthias and Johnson, Peter D. and Kieferov{\'{a}}, M{\'{a}}ria and Kivlichan, Ian D. and Menke, Tim and Peropadre, Borja and Sawaya, Nicolas P. D. and Sim, Sukin and Veis, Libor and Aspuru-Guzik, Al{\'{a}}n},
doi = {10.1021/acs.chemrev.8b00803},
issn = {0009-2665},
journal = {Chem. Rev.},
month = {oct},
number = {19},
pages = {10856--10915},
title = {{Quantum Chemistry in the Age of Quantum Computing}},
url = {https://pubs.acs.org/doi/10.1021/acs.chemrev.8b00803},
volume = {119},
year = {2019}
}

@article{Gidney2021,
author = {Gidney, Craig and Eker{\aa}, Martin},
doi = {10.22331/q-2021-04-15-433},
issn = {2521-327X},
journal = {Quantum},
month = {apr},
pages = {433},
title = {{How to factor 2048 bit RSA integers in 8 hours using 20 million noisy qubits}},
url = {https://quantum-journal.org/papers/q-2021-04-15-433/},
volume = {5},
year = {2021}
}

@article{reiher2017elucidating,
  title={Elucidating reaction mechanisms on quantum computers},
  author={Reiher, Markus and Wiebe, Nathan and Svore, Krysta M and Wecker, Dave and Troyer, Matthias},
  journal={Proceedings of the {N}ational {A}cademy of {S}ciences},
  volume={114},
  number={29},
  pages={7555--7560},
  year={2017},
  publisher={National Acad Sciences}
}

@article{li2019electronic,
  title={The electronic complexity of the ground-state of the FeMo cofactor of nitrogenase as relevant to quantum simulations},
  author={Li, Zhendong and Li, Junhao and Dattani, Nikesh S and Umrigar, CJ and Chan, Garnet Kin-Lic},
  journal={The Journal of Chemical Physics},
  volume={150},
  number={2},
  pages={024302},
  year={2019},
  publisher={AIP Publishing LLC}
}

@inproceedings{raffenettiMPI2017,
author = {Raffenetti, Ken and Amer, Abdelhalim and Oden, Lena and Archer, Charles and Bland, Wesley and Fujita, Hajime and Guo, Yanfei and Janjusic, Tomislav and Durnov, Dmitry and Blocksome, Michael and Si, Min and Seo, Sangmin and Langer, Akhil and Zheng, Gengbin and Takagi, Masamichi and Coffman, Paul and Jose, Jithin and Sur, Sayantan and Sannikov, Alexander and Oblomov, Sergey and Chuvelev, Michael and Hatanaka, Masayuki and Zhao, Xin and Fischer, Paul and Rathnayake, Thilina and Otten, Matt and Min, Misun and Balaji, Pavan},
title = {Why is MPI so slow? analyzing the fundamental limits in implementing MPI-3.1},
year = {2017},
isbn = {9781450351140},
publisher = {Association for Computing Machinery},
address = {New York, NY, USA},
url = {https://doi.org/10.1145/3126908.3126963},
doi = {10.1145/3126908.3126963},
booktitle = {Proceedings of the International Conference for High Performance Computing, Networking, Storage and Analysis},
articleno = {62},
numpages = {12},
location = {Denver, Colorado},
series = {SC '17}
}

@article{otten2016mpi,
  title={An MPI/OpenACC implementation of a high-order electromagnetics solver with GPUDirect communication},
  author={Otten, Matthew and Gong, Jing and Mametjanov, Azamat and Vose, Aaron and Levesque, John and Fischer, Paul and Min, Misun},
  journal={The International Journal of High Performance Computing Applications},
  volume={30},
  number={3},
  pages={320--334},
  year={2016},
  publisher={Sage Publications Sage UK: London, England}
}

@article{mitra2024localized,
  title={The localized active space method with unitary selective coupled cluster},
  author={Mitra, Abhishek and D’Cunha, Ruhee and Wang, Qiaohong and Hermes, Matthew R and Alexeev, Yuri and Gray, Stephen K and Otten, Matthew and Gagliardi, Laura},
  journal={Journal of chemical theory and computation},
  volume={20},
  number={18},
  pages={7865--7875},
  year={2024},
  publisher={ACS Publications}
}

@article{fedorov2022unitary,
  title={Unitary selective coupled-cluster method},
  author={Fedorov, Dmitry A and Alexeev, Yuri and Gray, Stephen K and Otten, Matthew},
  journal={Quantum},
  volume={6},
  pages={703},
  year={2022},
  publisher={Verein zur F{\"o}rderung des Open Access Publizierens in den Quantenwissenschaften}
}

@article{watts2024fullerene,
  title={Fullerene-encapsulated Cyclic Ozone for the Next Generation of Nano-sized Propellants via Quantum Computation},
  author={Watts, Thomas W and Otten, Matthew and Necaise, Jason T and Nguyen, Nam and Link, Benjamin and Williams, Kristen S and Sanders, Yuval R and Elman, Samuel J and Kieferova, Maria and Bremner, Michael J and others},
  journal={arXiv:2408.13244},
  year={2024},
  month={08}
}

@article{otten2024quantum,
  title={Quantum Resources Required for Binding Affinity Calculations of Amyloid beta},
  author={Otten, Matthew and Watts, Thomas W and Johnson, Samuel D and Sundareswara, Rashmi and Wang, Zhihui and Hardikar, Tarini S and Heitritter, Kenneth and Brown, James and Setia, Kanav and Holmes, Adam},
  url={https://arxiv.org/abs/2406.18744},
  journal={arXiv:2406.18744},
  month={06},
  year={2024}
}

@article{nguyen2024quantum,
  title={Quantum computing for corrosion-resistant materials and anti-corrosive coatings design},
  author={Nguyen, Nam and Watts, Thomas W and Link, Benjamin and Williams, Kristen S and Sanders, Yuval R and Elman, Samuel J and Kieferova, Maria and Bremner, Michael J and Morrell, Kaitlyn J and Elenewski, Justin and others},
  url={https://arxiv.org/abs/2406.18759},
  journal={arXiv:2406.18759},
  month={06},
  year={2024}
}

@article{Pattison2024,
author = {Pattison, Christopher A. and Baranes, Gefen and Ataides, J. Pablo Bonilla and Lukin, Mikhail D. and Zhou, Hengyun},
journal = {arXiv Prepr.},
month = {aug},
title = {{Fast quantum interconnects via constant-rate entanglement distillation}},
url = {http://arxiv.org/abs/2408.15936},
volume = {2408.15936},
year = {2024}
}

@article{Hartung2024,
author = {Hartung, Lukas and Seubert, Matthias and Welte, Stephan and Distante, Emanuele and Rempe, Gerhard},
doi = {10.1126/science.ado6471},
issn = {0036-8075},
journal = {Science},
month = {jul},
number = {6705},
pages = {179--183},
title = {{A quantum-network register assembled with optical tweezers in an optical cavity}},
url = {https://www.science.org/doi/10.1126/science.ado6471},
volume = {385},
year = {2024}
}

@article{Li2025,
archivePrefix = {arXiv},
author = {Li, Lintao and Hu, Xiye and Jia, Zhubing and Huie, William and Sun, Won Kyu Calvin and Aakash and Dong, Yuhao and Hiri-O-Tuppa, Narisak and Covey, Jacob P.},
journal = {arXiv Prepr.},
month = {feb},
title = {{Parallelized telecom quantum networking with a ytterbium-171 atom array}},
url = {http://arxiv.org/abs/2502.17406},
volume = {2502.17406},
year = {2025}
}

@article{Wehner2018,
author = {Wehner, Stephanie and Elkouss, David and Hanson, Ronald},
doi = {10.1126/science.aam9288},
issn = {0036-8075},
journal = {Science (80-. ).},
month = {oct},
number = {6412},
pages = {eaam9288},
title = {{Quantum internet: A vision for the road ahead}},
url = {https://www.sciencemag.org/lookup/doi/10.1126/science.aam9288},
volume = {362},
year = {2018}
}

@article{Erhard2021,
author = {Erhard, Alexander and {Poulsen Nautrup}, Hendrik and Meth, Michael and Postler, Lukas and Stricker, Roman and Stadler, Martin and Negnevitsky, Vlad and Ringbauer, Martin and Schindler, Philipp and Briegel, Hans J. and Blatt, Rainer and Friis, Nicolai and Monz, Thomas},
doi = {10.1038/s41586-020-03079-6},
issn = {0028-0836},
journal = {Nature},
month = {jan},
number = {7841},
pages = {220--224},
title = {{Entangling logical qubits with lattice surgery}},
url = {http://www.nature.com/articles/s41586-020-03079-6},
volume = {589},
year = {2021}
}

@article{Wan2019,
author = {Wan, Yong and Kienzler, Daniel and Erickson, Stephen D. and Mayer, Karl H. and Tan, Ting Rei and Wu, Jenny J. and Vasconcelos, Hilma M. and Glancy, Scott and Knill, Emanuel and Wineland, David J. and Wilson, Andrew C. and Leibfried, Dietrich},
doi = {10.1126/science.aaw9415},
issn = {0036-8075},
journal = {Science (80-. ).},
month = {may},
number = {6443},
pages = {875--878},
title = {{Quantum gate teleportation between separated qubits in a trapped-ion processor}},
url = {https://www.sciencemag.org/lookup/doi/10.1126/science.aaw9415},
volume = {364},
year = {2019}
}

@article{vanLeent2022,
author = {van Leent, Tim and Bock, Matthias and Fertig, Florian and Garthoff, Robert and Eppelt, Sebastian and Zhou, Yiru and Malik, Pooja and Seubert, Matthias and Bauer, Tobias and Rosenfeld, Wenjamin and Zhang, Wei and Becher, Christoph and Weinfurter, Harald},
doi = {10.1038/s41586-022-04764-4},
issn = {0028-0836},
journal = {Nature},
month = {jul},
number = {7917},
pages = {69--73},
title = {{Entangling single atoms over 33 km telecom fibre}},
url = {https://www.nature.com/articles/s41586-022-04764-4},
volume = {607},
year = {2022}
}

@article{DCunha2024,
author = {D'Cunha, Ruhee and Otten, Matthew and Hermes, Matthew R. and Gagliardi, Laura and Gray, Stephen K.},
doi = {10.1021/acs.jctc.3c01283},
issn = {1549-9618},
journal = {J. Chem. Theory Comput.},
month = {apr},
number = {8},
pages = {3121--3130},
title = {{State Preparation in Quantum Algorithms for Fragment-Based Quantum Chemistry}},
url = {https://pubs.acs.org/doi/10.1021/acs.jctc.3c01283},
volume = {20},
year = {2024}
}

@article{Bluvstein2023,
author = {Bluvstein, Dolev and Evered, Simon J. and Geim, Alexandra A. and Li, Sophie H. and Zhou, Hengyun and Manovitz, Tom and Ebadi, Sepehr and Cain, Madelyn and Kalinowski, Marcin and Hangleiter, Dominik and Ataides, J. Pablo Bonilla and Maskara, Nishad and Cong, Iris and Gao, Xun and Rodriguez, Pedro Sales and Karolyshyn, Thomas and Semeghini, Giulia and Gullans, Michael J. and Greiner, Markus and Vuleti{\'{c}}, Vladan and Lukin, Mikhail D.},
doi = {10.1038/s41586-023-06927-3},
issn = {0028-0836},
journal = {Nature},
month = {dec},
title = {{Logical quantum processor based on reconfigurable atom arrays}},
url = {https://www.nature.com/articles/s41586-023-06927-3},
year = {2023}
}

@article{Manetsch2024,
  title={A tweezer array with 6100 highly coherent atomic qubits},
  author={Manetsch, Hannah J and Nomura, Gyohei and Bataille, Elie and Lv, Xudong and Leung, Kon H and Endres, Manuel},
  journal={Nature},
  pages={1--3},
  year={2025},
  publisher={Nature Publishing Group UK London}
}

@article{doi:10.1021/acs.jpca.8b01554,
author = {Chien, Alan D. and Holmes, Adam A. and Otten, Matthew and Umrigar, C. J. and Sharma, Sandeep and Zimmerman, Paul M.},
title = {Excited States of Methylene, Polyenes, and Ozone from Heat-Bath Configuration Interaction},
journal = {The Journal of Physical Chemistry A},
volume = {122},
number = {10},
pages = {2714-2722},
year = {2018},
doi = {10.1021/acs.jpca.8b01554},
    note ={PMID: 29473750},

URL = { 
    
        https://doi.org/10.1021/acs.jpca.8b01554
    
    

},
eprint = { 
    
        https://doi.org/10.1021/acs.jpca.8b01554
    
    

}

}

@article{bellonzi2024feasibility,
  title={Feasibility of accelerating homogeneous catalyst discovery with fault-tolerant quantum computers},
  author={Nicole Bellonzi and Alexander Kunitsa and Joshua T. Cantin and Jorge A. Campos-Gonzalez-Angulo and Maxwell D. Radin and Yanbing Zhou and Peter D. Johnson and Luis A. Martínez-Martínez and Mohammad Reza Jangrouei and Aritra Sankar Brahmachari and Linjun Wang and Smik Patel and Monika Kodrycka and Ignacio Loaiza and Robert A. Lang and Alán Aspuru-Guzik and Artur F. Izmaylov and Jhonathan Romero Fontalvo and Yudong Cao},
  journal={arXiv preprint arXiv:2406.06335},
  year={2024}
}

@article{sawaya2022constructing,
  title={Constructing Hubbard models for the hydrogen chain using sliced-basis density matrix renormalization group},
  author={Sawaya, Randy C and White, Steven R},
  journal={Physical Review B},
  volume={105},
  number={4},
  pages={045145},
  year={2022},
  publisher={APS}
}

@article{PhysRevX.7.031059,
  title = {Towards the Solution of the Many-Electron Problem in Real Materials: Equation of State of the Hydrogen Chain with State-of-the-Art Many-Body Methods},
  author = {Motta, Mario and Ceperley, David M. and Chan, Garnet Kin-Lic and Gomez, John A. and Gull, Emanuel and Guo, Sheng and Jim\'enez-Hoyos, Carlos A. and Lan, Tran Nguyen and Li, Jia and Ma, Fengjie and Millis, Andrew J. and Prokof'ev, Nikolay V. and Ray, Ushnish and Scuseria, Gustavo E. and Sorella, Sandro and Stoudenmire, Edwin M. and Sun, Qiming and Tupitsyn, Igor S. and White, Steven R. and Zgid, Dominika and Zhang, Shiwei},
  collaboration = {Simons Collaboration on the Many-Electron Problem},
  journal = {Phys. Rev. X},
  volume = {7},
  issue = {3},
  pages = {031059},
  numpages = {28},
  year = {2017},
  month = {Sep},
  publisher = {American Physical Society},
  doi = {10.1103/PhysRevX.7.031059},
  url = {https://link.aps.org/doi/10.1103/PhysRevX.7.031059}
}

@article{low2019hamiltonian,
  title={Hamiltonian simulation by qubitization},
  author={Low, Guang Hao and Chuang, Isaac L},
  journal={Quantum},
  volume={3},
  pages={163},
  year={2019},
  publisher={Verein zur F{\"o}rderung des Open Access Publizierens in den Quantenwissenschaften}
}

@article{lee2021even,
  title={Even more efficient quantum computations of chemistry through tensor hypercontraction},
  author={Lee, Joonho and Berry, Dominic W and Gidney, Craig and Huggins, William J and McClean, Jarrod R and Wiebe, Nathan and Babbush, Ryan},
  journal={PRX Quantum},
  volume={2},
  number={3},
  pages={030305},
  year={2021},
  publisher={APS}
}

@article{otten2023qrechem,
  title={QREChem: quantum resource estimation software for chemistry applications},
  author={Otten, Matthew and Kang, Byeol and Fedorov, Dmitry and Lee, Joo-Hyoung and Benali, Anouar and Habib, Salman and Gray, Stephen K and Alexeev, Yuri},
  journal={Frontiers in Quantum Science and Technology},
  volume={2},
  pages={1232624},
  year={2023},
  publisher={Frontiers Media SA}
}

@article{low2025fast,
  title={Fast quantum simulation of electronic structure by spectrum amplification},
  author={Low, Guang Hao and King, Robbie and Berry, Dominic W and Han, Qiushi and DePrince III, A Eugene and White, Alec and Babbush, Ryan and Somma, Rolando D and Rubin, Nicholas C},
  journal={arXiv preprint arXiv:2502.15882},
  year={2025}
}

@article{mohseni2024build,
  title={How to build a quantum supercomputer: Scaling challenges and opportunities},
  author={Masoud Mohseni and Artur Scherer and K. Grace Johnson and Oded Wertheim and Matthew Otten and Navid Anjum Aadit and Yuri Alexeev and Kirk M. Bresniker and Kerem Y. Camsari and Barbara Chapman and Soumitra Chatterjee and Gebremedhin A. Dagnew and Aniello Esposito and Farah Fahim and Marco Fiorentino and Archit Gajjar and Abdullah Khalid and Xiangzhou Kong and Bohdan Kulchytskyy and Elica Kyoseva and Ruoyu Li and P. Aaron Lott and Igor L. Markov and Robert F. McDermott and Giacomo Pedretti and Pooja Rao and Eleanor Rieffel and Allyson Silva and John Sorebo and Panagiotis Spentzouris and Ziv Steiner and Boyan Torosov and Davide Venturelli and Robert J. Visser and Zak Webb and Xin Zhan and Yonatan Cohen and Pooya Ronagh and Alan Ho and Raymond G. Beausoleil and John M. Martinis},
  journal={arXiv preprint arXiv:2411.10406},
  year={2024},
}

@article{Otten2022,
author = {Otten, Matthew and Hermes, Matthew R. and Pandharkar, Riddhish and Alexeev, Yuri and Gray, Stephen K. and Gagliardi, Laura},
doi = {10.1021/acs.jctc.2c00388},
issn = {1549-9618},
journal = {J. Chem. Theory Comput.},
month = {dec},
number = {12},
pages = {7205--7217},
title = {{Localized Quantum Chemistry on Quantum Computers}},
url = {https://pubs.acs.org/doi/10.1021/acs.jctc.2c00388},
volume = {18},
year = {2022}
}

@article{Monroe2014,
author = {Monroe, C. and Raussendorf, R. and Ruthven, A. and Brown, K. R. and Maunz, P. and Duan, L.-M. and Kim, J.},
doi = {10.1103/PhysRevA.89.022317},
issn = {1050-2947},
journal = {Phys. Rev. A},
month = {feb},
number = {2},
pages = {022317},
title = {{Large-scale modular quantum-computer architecture with atomic memory and photonic interconnects}},
url = {https://link.aps.org/doi/10.1103/PhysRevA.89.022317},
volume = {89},
year = {2014}
}

@article{Beverland2022,
archivePrefix = {arXiv},
arxivId = {2211.07629},
author = {Beverland, Michael E. and Murali, Prakash and Troyer, Matthias and Svore, Krysta M. and Hoefler, Torsten and Kliuchnikov, Vadym and Low, Guang Hao and Soeken, Mathias and Sundaram, Aarthi and Vaschillo, Alexander},
eprint = {2211.07629},
journal = {arXiv Prepr.},
month = {nov},
title = {{Assessing requirements to scale to practical quantum advantage}},
url = {http://arxiv.org/abs/2211.07629},
volume = {2211.07629},
year = {2022}
}

@article{Bluvstein2022,
author = {Bluvstein, Dolev and Levine, Harry and Semeghini, Giulia and Wang, Tout T. and Ebadi, Sepehr and Kalinowski, Marcin and Keesling, Alexander and Maskara, Nishad and Pichler, Hannes and Greiner, Markus and Vuleti{\'{c}}, Vladan and Lukin, Mikhail D.},
doi = {10.1038/s41586-022-04592-6},
issn = {0028-0836},
journal = {Nature},
month = {apr},
number = {7906},
pages = {451--456},
title = {{A quantum processor based on coherent transport of entangled atom arrays}},
url = {https://www.nature.com/articles/s41586-022-04592-6},
volume = {604},
year = {2022}
}

@article{Covey2023,
author = {Covey, Jacob P. and Weinfurter, Harald and Bernien, Hannes},
doi = {10.1038/s41534-023-00759-9},
issn = {2056-6387},
journal = {npj Quantum Inf.},
month = {sep},
number = {1},
pages = {90},
title = {{Quantum networks with neutral atom processing nodes}},
url = {https://www.nature.com/articles/s41534-023-00759-9},
volume = {9},
year = {2023}
}

@article{LiThompson2024,
author = {Li, Yiyi and Thompson, Jeff D.},
doi = {10.1103/PRXQuantum.5.020363},
issn = {2691-3399},
journal = {PRX Quantum},
month = {jun},
number = {2},
pages = {020363},
title = {{High-Rate and High-Fidelity Modular Interconnects between Neutral Atom Quantum Processors}},
url = {https://link.aps.org/doi/10.1103/PRXQuantum.5.020363},
volume = {5},
year = {2024}
}

@article{Ruhee2024,
author = {D’Cunha, Ruhee and Otten, Matthew and Hermes, Matthew R. and Gagliardi, Laura and Gray, Stephen K.},
title = {State Preparation in Quantum Algorithms for Fragment-Based Quantum Chemistry},
journal = {Journal of Chemical Theory and Computation},
volume = {20},
number = {8},
pages = {3121-3130},
year = {2024},
doi = {10.1021/acs.jctc.3c01283},
    note ={PMID: 38607377},

URL = { 
    
        https://doi.org/10.1021/acs.jctc.3c01283
    
    

},
eprint = { 
    
        https://doi.org/10.1021/acs.jctc.3c01283
    
    

}
}

@article{Verma2025,
  title = {Polynomial Scaling Localized Active Space Unitary Selective Coupled Cluster Singles and Doubles},
  author = {Verma, S. and D’Cunha, R. and Mitra, A. and Hermes, M. R. and Gray, S. K. and Otten, M. and others},
  journal = {ChemRxiv},
  year = {2025},
  note = {This content is a preprint and has not been peer-reviewed},
  doi = {10.26434/chemrxiv-2025-1n3d4-v2}
}

@INPROCEEDINGS{Dillon2010,
  author={Dillon, Tharam and Wu, Chen and Chang, Elizabeth},
  booktitle={2010 24th IEEE International Conference on Advanced Information Networking and Applications}, 
  title={Cloud Computing: Issues and Challenges}, 
  year={2010},
  volume={},
  number={},
  pages={27-33},
  doi={10.1109/AINA.2010.187}}

@ARTICLE{OpenMP1998,
  author={Dagum, L. and Menon, R.},
  journal={IEEE Computational Science and Engineering}, 
  title={OpenMP: an industry standard API for shared-memory programming}, 
  year={1998},
  volume={5},
  number={1},
  pages={46-55},
  doi={10.1109/99.660313}}

@article{SSH1979,
  title = {Solitons in Polyacetylene},
  author = {Su, W. P. and Schrieffer, J. R. and Heeger, A. J.},
  journal = {Phys. Rev. Lett.},
  volume = {42},
  issue = {25},
  pages = {1698--1701},
  numpages = {0},
  year = {1979},
  month = {Jun},
  publisher = {American Physical Society},
  doi = {10.1103/PhysRevLett.42.1698},
  url = {https://link.aps.org/doi/10.1103/PhysRevLett.42.1698}
}

@article{Catarina2023,
	title = {Density-matrix renormalization group: a pedagogical introduction},
	volume = {96},
	issn = {1434-6036},
	url = {https://doi.org/10.1140/epjb/s10051-023-00575-2},
	doi = {10.1140/epjb/s10051-023-00575-2},
	number = {8},
	journal = {The European Physical Journal B},
	author = {Catarina, G. and Murta, Bruno},
	month = aug,
	year = {2023},
	pages = {111},
}

@techreport{bianco2013,
  title        = {An Interface for Halo Exchange Pattern},
  author       = {Mauro Bianco},
  year         = {2013},
  institution  = {Partnership for Advanced Computing in Europe (PRACE)},
  number       = {WP86},
  url          = {https://prace-ri.eu/wp-content/uploads/wp86.pdf},
  note         = {PRACE White Paper}
}

@article{Barkoutsos2018,
   title={Quantum algorithms for electronic structure calculations: Particle-hole Hamiltonian and optimized wave-function expansions},
   volume={98},
   ISSN={2469-9934},
   url={http://dx.doi.org/10.1103/PhysRevA.98.022322},
   DOI={10.1103/physreva.98.022322},
   number={2},
   journal={Physical Review A},
   publisher={American Physical Society (APS)},
   author={Barkoutsos, Panagiotis Kl. and Gonthier, Jerome F. and Sokolov, Igor and Moll, Nikolaj and Salis, Gian and Fuhrer, Andreas and Ganzhorn, Marc and Egger, Daniel J. and Troyer, Matthias and Mezzacapo, Antonio and Filipp, Stefan and Tavernelli, Ivano},
   year={2018},
   month=aug }

@article{Horsman2012,
   title={Surface code quantum computing by lattice surgery},
   volume={14},
   ISSN={1367-2630},
   url={http://dx.doi.org/10.1088/1367-2630/14/12/123011},
   DOI={10.1088/1367-2630/14/12/123011},
   number={12},
   journal={New Journal of Physics},
   publisher={IOP Publishing},
   author={Horsman, Dominic and Fowler, Austin G and Devitt, Simon and Meter, Rodney Van},
   year={2012},
   month=dec, pages={123011} }

@article{Childs2021,
   title={Theory of Trotter Error with Commutator Scaling},
   volume={11},
   ISSN={2160-3308},
   url={http://dx.doi.org/10.1103/PhysRevX.11.011020},
   DOI={10.1103/physrevx.11.011020},
   number={1},
   journal={Physical Review X},
   publisher={American Physical Society (APS)},
   author={Childs, Andrew M. and Su, Yuan and Tran, Minh C. and Wiebe, Nathan and Zhu, Shuchen},
   year={2021},
   month=feb }

@article{pople1965,
  title     = {Self‐Consistent Molecular‐Orbital Methods. I. Use of Gaussian Expansions of Slater‐Type Atomic Orbitals},
  author    = {Pople, John A. and Hehre, Warren J.},
  journal   = {The Journal of Chemical Physics},
  volume    = {43},
  number    = {10},
  pages     = {S229--S236},
  year      = {1965},
  publisher = {American Institute of Physics},
  doi       = {10.1063/1.1672392}
}

@misc{Ovrum2007,
      title={Quantum computation algorithm for many-body studies}, 
      author={E. Ovrum and M. Hjorth-Jensen},
      year={2007},
      eprint={0705.1928},
      archivePrefix={arXiv},
      primaryClass={quant-ph},
      url={https://arxiv.org/abs/0705.1928}, 
}

@article{Berry2006,
   title={Efficient Quantum Algorithms for Simulating Sparse Hamiltonians},
   volume={270},
   ISSN={1432-0916},
   url={http://dx.doi.org/10.1007/s00220-006-0150-x},
   DOI={10.1007/s00220-006-0150-x},
   number={2},
   journal={Communications in Mathematical Physics},
   publisher={Springer Science and Business Media LLC},
   author={Berry, Dominic W. and Ahokas, Graeme and Cleve, Richard and Sanders, Barry C.},
   year={2006},
   month=dec, pages={359–371} }

@article{Ito2025,
  title = {Efficient Magic State Distillation by Zero-level Distillation},
    author = {Itogawa, Tomohiro and Takada, Yugo and Hirano, Yutaka and Fujii, Keisuke},
  journal = {PRX Quantum},
  pages = {--},
  year = {2025},
  month = {May},
  publisher = {American Physical Society},
  doi = {10.1103/thxx-njr6},
  url = {https://link.aps.org/doi/10.1103/thxx-njr6}
}

@article{Akahoshi2024,
  title = {Partially Fault-Tolerant Quantum Computing Architecture with Error-Corrected Clifford Gates and Space-Time Efficient Analog Rotations},
  author = {Akahoshi, Yutaro and Maruyama, Kazunori and Oshima, Hirotaka and Sato, Shintaro and Fujii, Keisuke},
  journal = {PRX Quantum},
  volume = {5},
  issue = {1},
  pages = {010337},
  numpages = {21},
  year = {2024},
  month = {Mar},
  publisher = {American Physical Society},
  doi = {10.1103/PRXQuantum.5.010337},
  url = {https://link.aps.org/doi/10.1103/PRXQuantum.5.010337}
}

@misc{Gidney2025,
      title={How to factor 2048 bit RSA integers with less than a million noisy qubits}, 
      author={Craig Gidney},
      year={2025},
      eprint={2505.15917},
      archivePrefix={arXiv},
      primaryClass={quant-ph},
      url={https://arxiv.org/abs/2505.15917}, 
}

@misc{Zhou2025,
      title={Resource Analysis of Low-Overhead Transversal Architectures for Reconfigurable Atom Arrays}, 
      author={Hengyun Zhou and Casey Duckering and Chen Zhao and Dolev Bluvstein and Madelyn Cain and Aleksander Kubica and Sheng-Tao Wang and Mikhail D. Lukin},
      year={2025},
      eprint={2505.15907},
      archivePrefix={arXiv},
      primaryClass={quant-ph},
      url={https://arxiv.org/abs/2505.15907}, 
}

@article{Hermes2020,
	title = {Variational {Localized} {Active} {Space} {Self}-{Consistent} {Field} {Method}},
	volume = {16},
	issn = {1549-9618},
	url = {https://doi.org/10.1021/acs.jctc.0c00222},
	doi = {10.1021/acs.jctc.0c00222},
	number = {8},
	journal = {Journal of Chemical Theory and Computation},
	author = {Hermes, Matthew R. and Pandharkar, Riddhish and Gagliardi, Laura},
	month = aug,
	year = {2020},
	note = {Publisher: American Chemical Society},
	pages = {4923--4937},
}

@misc{TianRepo,
  author       = {Tian Xue},
  title        = {H-chain},
  year         = {2025},
  howpublished = {\url{https://github.com/TianXue2002/H-chain}},
  note         = {Accessed: 2025-06-15}
}

@article{Cowtan2020,
   title={Phase Gadget Synthesis for Shallow Circuits},
   volume={318},
   ISSN={2075-2180},
   url={http://dx.doi.org/10.4204/EPTCS.318.13},
   DOI={10.4204/eptcs.318.13},
   journal={Electronic Proceedings in Theoretical Computer Science},
   publisher={Open Publishing Association},
   author={Cowtan, Alexander and Dilkes, Silas and Duncan, Ross and Simmons, Will and Sivarajah, Seyon},
   year={2020},
   month=may, pages={213–228} }

@article{Chamberland2018,
   title={Fault-tolerant quantum computing in the Pauli or Clifford frame with slow error diagnostics},
   volume={2},
   ISSN={2521-327X},
   url={http://dx.doi.org/10.22331/q-2018-01-04-43},
   DOI={10.22331/q-2018-01-04-43},
   journal={Quantum},
   publisher={Verein zur Forderung des Open Access Publizierens in den Quantenwissenschaften},
   author={Chamberland, Christopher and Iyer, Pavithran and Poulin, David},
   year={2018},
   month=jan, pages={43} }

@misc{Chen2024,
      title={Transversal Logical Clifford gates on rotated surface codes with reconfigurable neutral atom arrays}, 
      author={Zi-Han Chen and Ming-Cheng Chen and Chao-Yang Lu and Jian-Wei Pan},
      year={2024},
      eprint={2412.01391},
      archivePrefix={arXiv},
      primaryClass={quant-ph},
      url={https://arxiv.org/abs/2412.01391}, 
}

@misc{Ross2016,
      title={Optimal ancilla-free Clifford+T approximation of z-rotations}, 
      author={Neil J. Ross and Peter Selinger},
      year={2016},
      eprint={1403.2975},
      archivePrefix={arXiv},
      primaryClass={quant-ph},
      url={https://arxiv.org/abs/1403.2975}, 
}

@misc{Gidney2024,
      title={Magic state cultivation: growing T states as cheap as CNOT gates}, 
      author={Craig Gidney and Noah Shutty and Cody Jones},
      year={2024},
      eprint={2409.17595},
      archivePrefix={arXiv},
      primaryClass={quant-ph},
      url={https://arxiv.org/abs/2409.17595}, 
}

@misc{Chen2024RP2,
      title={Transversal Logical Clifford gates on rotated surface codes with reconfigurable neutral atom arrays}, 
      author={Zi-Han Chen and Ming-Cheng Chen and Chao-Yang Lu and Jian-Wei Pan},
      year={2024},
      eprint={2412.01391},
      archivePrefix={arXiv},
      primaryClass={quant-ph},
      url={https://arxiv.org/abs/2412.01391}, 
}

@article{Shreya2025,
author = {Verma, Shreya and D’Cunha, Ruhee and Mitra, Abhishek and Hermes, Matthew and Gray, Stephen K. and Otten, Matthew and Gagliardi, Laura},
title = {Polynomial Scaling Localized Active Space Unitary Selective Coupled Cluster Singles and Doubles},
journal = {Journal of Chemical Theory and Computation},
volume = {21},
number = {15},
pages = {7460-7470},
year = {2025},
doi = {10.1021/acs.jctc.5c00745},
    note ={PMID: 40669079},

URL = { 
https://doi.org/10.1021/acs.jctc.5c00745},
eprint = { 
https://doi.org/10.1021/acs.jctc.5c00745}
}

@article{Ouyang2020,
   title={Compilation by stochastic Hamiltonian sparsification},
   volume={4},
   ISSN={2521-327X},
   url={http://dx.doi.org/10.22331/q-2020-02-27-235},
   DOI={10.22331/q-2020-02-27-235},
   journal={Quantum},
   publisher={Verein zur Forderung des Open Access Publizierens in den Quantenwissenschaften},
   author={Ouyang, Yingkai and White, David R. and Campbell, Earl T.},
   year={2020},
   month=feb, pages={235} }

@article{Childs2019,
   title={Faster quantum simulation by randomization},
   volume={3},
   ISSN={2521-327X},
   url={http://dx.doi.org/10.22331/q-2019-09-02-182},
   DOI={10.22331/q-2019-09-02-182},
   journal={Quantum},
   publisher={Verein zur Forderung des Open Access Publizierens in den Quantenwissenschaften},
   author={Childs, Andrew M. and Ostrander, Aaron and Su, Yuan},
   year={2019},
   month=sep, pages={182} }

@article{Cambell2019,
  title = {Random Compiler for Fast Hamiltonian Simulation},
  author = {Campbell, Earl},
  journal = {Phys. Rev. Lett.},
  volume = {123},
  issue = {7},
  pages = {070503},
  numpages = {5},
  year = {2019},
  month = {Aug},
  publisher = {American Physical Society},
  doi = {10.1103/PhysRevLett.123.070503},
  url = {https://link.aps.org/doi/10.1103/PhysRevLett.123.070503}
}

@misc{Gidney2019,
      title={Approximate encoded permutations and piecewise quantum adders}, 
      author={Craig Gidney},
      year={2019},
      eprint={1905.08488},
      archivePrefix={arXiv},
      primaryClass={quant-ph},
      url={https://arxiv.org/abs/1905.08488}, 
}

@misc{Sahay2025,
      title={Fold-transversal surface code cultivation}, 
      author={Kaavya Sahay and Pei-Kai Tsai and Kathleen Chang and Qile Su and Thomas B. Smith and Shraddha Singh and Shruti Puri},
      year={2025},
      eprint={2509.05212},
      archivePrefix={arXiv},
      primaryClass={quant-ph},
      url={https://arxiv.org/abs/2509.05212}, 
}

@article{Niu2023,
   title={Low-loss interconnects for modular superconducting quantum processors},
   volume={6},
   ISSN={2520-1131},
   url={http://dx.doi.org/10.1038/s41928-023-00925-z},
   DOI={10.1038/s41928-023-00925-z},
   number={3},
   journal={Nature Electronics},
   publisher={Springer Science and Business Media LLC},
   author={Niu, Jingjing and Zhang, Libo and Liu, Yang and Qiu, Jiawei and Huang, Wenhui and Huang, Jiaxiang and Jia, Hao and Liu, Jiawei and Tao, Ziyu and Wei, Weiwei and Zhou, Yuxuan and Zou, Wanjing and Chen, Yuanzhen and Deng, Xiaowei and Deng, Xiuhao and Hu, Changkang and Hu, Ling and Li, Jian and Tan, Dian and Xu, Yuan and Yan, Fei and Yan, Tongxing and Liu, Song and Zhong, Youpeng and Cleland, Andrew N. and Yu, Dapeng},
   year={2023},
   month=feb, pages={235–241} }

@misc{Wu2025,
      title={State Similarity in Modular Superconducting Quantum Processors with Classical Communications}, 
      author={Bujiao Wu and Changrong Xie and Peng Mi and Zhiyi Wu and Zechen Guo and Peisheng Huang and Wenhui Huang and Xuandong Sun and Jiawei Zhang and Libo Zhang and Jiawei Qiu and Xiayu Linpeng and Ziyu Tao and Ji Chu and Ji Jiang and Song Liu and Jingjing Niu and Yuxuan Zhou and Yuxuan Du and Wenhui Ren and Youpeng Zhong and Tongliang Liu and Dapeng Yu},
      year={2025},
      eprint={2506.01657},
      archivePrefix={arXiv},
      primaryClass={quant-ph},
      url={https://arxiv.org/abs/2506.01657}, 
}

@article{Ramette2024,
	title = {Fault-tolerant connection of error-corrected qubits with noisy links},
	volume = {10},
	issn = {2056-6387},
	url = {https://doi.org/10.1038/s41534-024-00855-4},
	doi = {10.1038/s41534-024-00855-4},
	number = {1},
	journal = {npj Quantum Information},
	author = {Ramette, Joshua and Sinclair, Josiah and Breuckmann, Nikolas P. and Vuletić, Vladan},
	month = jun,
	year = {2024},
	pages = {58},
}

@article{Vazquez2024,
   title={Combining quantum processors with real-time classical communication},
   volume={636},
   ISSN={1476-4687},
   url={http://dx.doi.org/10.1038/s41586-024-08178-2},
   DOI={10.1038/s41586-024-08178-2},
   number={8041},
   journal={Nature},
   publisher={Springer Science and Business Media LLC},
   author={Carrera Vazquez, Almudena and Tornow, Caroline and Ristè, Diego and Woerner, Stefan and Takita, Maika and Egger, Daniel J.},
   year={2024},
   month=nov, pages={75–79} }

@article{Babbush2018,
   title={Encoding Electronic Spectra in Quantum Circuits with Linear T Complexity},
   volume={8},
   ISSN={2160-3308},
   url={http://dx.doi.org/10.1103/PhysRevX.8.041015},
   DOI={10.1103/physrevx.8.041015},
   number={4},
   journal={Physical Review X},
   publisher={American Physical Society (APS)},
   author={Babbush, Ryan and Gidney, Craig and Berry, Dominic W. and Wiebe, Nathan and McClean, Jarrod and Paler, Alexandru and Fowler, Austin and Neven, Hartmut},
   year={2018},
   month=oct }

@misc{Low2025,
      title={Fast quantum simulation of electronic structure by spectrum amplification}, 
      author={Guang Hao Low and Robbie King and Dominic W. Berry and Qiushi Han and A. Eugene DePrince III and Alec White and Ryan Babbush and Rolando D. Somma and Nicholas C. Rubin},
      year={2025},
      eprint={2502.15882},
      archivePrefix={arXiv},
      primaryClass={quant-ph},
      url={https://arxiv.org/abs/2502.15882}, 
}

@article{Low2019,
  doi = {10.22331/q-2019-07-12-163},
  url = {https://doi.org/10.22331/q-2019-07-12-163},
  title = {Hamiltonian {S}imulation by {Q}ubitization},
  author = {Low, Guang Hao and Chuang, Isaac L.},
  journal = {{Quantum}},
  issn = {2521-327X},
  publisher = {{Verein zur F{\"{o}}rderung des Open Access Publizierens in den Quantenwissenschaften}},
  volume = {3},
  pages = {163},
  month = jul,
  year = {2019}
}

\setcounter{section}{0}



\appendix
\renewcommand\appendixname{APPENDIX}
\renewcommand\thesection{\Alph{section}}
\renewcommand\thesubsection{\arabic{subsection}}



\setcounter{figure}{0}
\renewcommand{\thefigure}{S\arabic{figure}}

\section{USCC loading}
\label{USCC loading}
\textbf{Unitary Selective Coupled-Cluster (USCC)}. USCC~\cite{Otten2022} is an efficient algorithm for quantum chemistry. \red{In USCC, the state is first prepared by the localized classical solutions followed by a VQE with USCC anstaz. USCC filters terms in the Hamiltonian with energy contribution lower than $\epsilon$ and feeds the ansatz to the VQE solver to approximate the ground state energy as illustrated in Fig. \ref{fig:schematic}(a).} The classical solution LASSCF (Localized Active Space Self-Consistent Field) wave function is an anti-symmetric product of $K$ localized wavefunctions~\cite{Hermes2020}:
\begin{equation}
    \ket{\Psi_{LAS}} = \prod_K \ket{\Psi_K} \otimes \ket{\Phi_D}
\end{equation}
$\ket{\Psi_K}$ is the active many-body localized wave functions, and $\ket{\Phi_D}$ is the filled orbital wave. Each localized active wave function $\Psi_K$ is loaded within one QPU using QPE circuits or direction initialization circuit found by Qiskit~\cite{Ruhee2024}. The USCC wave function enables interactions between localized clusters:
\begin{equation}
     \ket{\Psi_{LAS-USCC}(\textbf{t})} =\hat{U}_{USCC}(\textbf{t})\ket{\Psi_{QLAS}}
\end{equation}
where $\hat{U}_{USCC}$ is the inter-cluster parametrized ansatz:
\begin{align}
     \label{eq:USCC}
     \hat{U}_{USCC}(\textbf{t}) =& \exp\bigg[t_{ij}(a_i^\dagger a_j - h.c.) +\\
     &t_{ijkm}(a_i^\dagger a_j^\dagger a_k a_m - h.c.) \bigg]
\end{align}
The following VQE after the \red{state preparation} in Fig. \ref{fig:schematic}(a) finds the optimal hopping amplitudes $\textbf{t} = [t_{ij}, t_{ijkm}]$ that approximate the ground state energy.

In the USCC algorithm, only the hopping terms with contribution to the energy above the threshold $\epsilon$ are included,
\begin{equation}
    \left|\frac{\partial E_{USCC}}{ 2\partial t_\beta}\right|_{t_\beta = 0}\geq \epsilon.
\end{equation}
\red{dUSCC only keeps significant ansatz with energy gradients greater than $\epsilon$ to reduce the complexity of the algorithm.
On a high level, the Coulomb potential is the dominant potential in the Hamiltonian, so terms with greater hopping distance contribute less to the energy, and they are only selected at the lower $\epsilon$. Fig \ref{fig:schematic} (c) reflects this relation, as at the same $\epsilon$, a larger inter-module distance $d$ leads to a more delayed circuit since the energy of inter-module hoppings is higher and more inter-module terms are included in the dUSCC. Fig. \ref{fig:packing plot} further supports this hypothesis: as $\epsilon$ decreases, more inter-module and long-range tiles appear, suggesting dUSCC includes more inter-module hoppings at lower $\epsilon$.}

\textbf{USCC Circuit Loading}. 
In this section, we will provide a naive compilation of dUSCC by a combination of Trotterization and JW transformation. A more compact compilation using ZX calculus is in Appendix \ref{appendix: zx proof}. The loadable dUSCC is:
\begin{align}
    \hat{U}_{dUSCC} =& \prod_{ij}M_{ij}^\dagger\exp\left[t_{ij}\prod_k \sigma_k^z\right]M_{ij}+\\
    &\prod_{ij}M_{ijkm}^\dagger\exp\left[t_{ijkm}\prod_k \sigma_k^z\right]M_{ijkm}
\end{align}
where $M_{ij}, M_{ijkm}$ are unitary matrices $M\in\{H,Y\}$ and $\exp\left[-it/2\prod_k \sigma^z_k\right]$ is a $R_Z(t)$ conjugated by ladders of CNOTs.

The sum on the exponents in the USCC ansatz from Eq. (\ref{eq:USCC}) can be approximated by the product of individual terms by Trotterization:
\begin{align}
    \hat{U}_{USCC} = &\prod\limits_{ij}\exp(t_{ij}a_i^\dagger a_j - h.c.) + \\
    &\prod\limits_{ijkm}\exp(t_{ijkm}a_i^\dagger a_j^\dagger a_k a_m - h.c.) + O(t^2)
\end{align}
Each individual fermionic USCC term is loaded on the quantum circuit by the Jordan-Wigner transformation:
\begin{align}
     a_i &= \left(\prod\limits_{j=1}^{i-1}\sigma^z_j\right) Q_i^+ 
  & a_i^\dagger &= \left(\prod\limits_{j=1}^{i-1}\sigma^z_j\right)Q_i^-\\
 Q^+ &= \frac{1}{2}(\sigma^x -i\sigma^y) & Q^- &= \frac{1}{2}(\sigma^x +i\sigma^y)
\end{align}

\begin{figure}[t]
	\includegraphics[width=\linewidth]{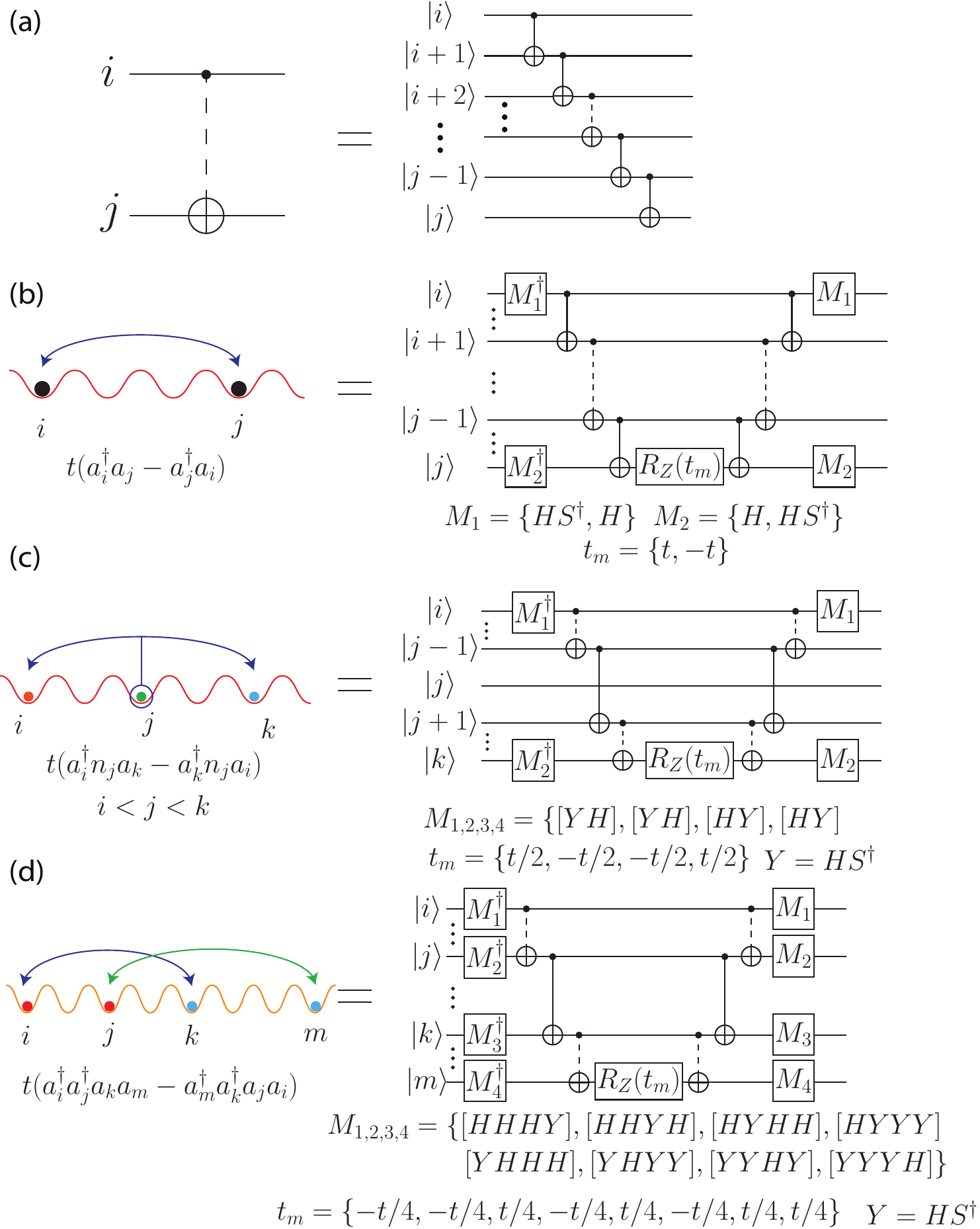}
	\caption{
        \textbf{The JW mapping}. (a) Each dashed CNOT represents a ladder of CNOTs. (b) The single hopping can be mapped to two conjugations of ladders of CNOTs, where the parameters are $M_1 = Y, M_2 = H, t = t_m$ and $M_1 = H, M_2 = HS, t = -t_m$ where $Y = HS^\dagger$. (c) The controlled hopping is mapped to four conjugations of ladders of CNOTs. We only show the case $i < j <k$ here. The circuit will be different at other conditions. (d) The double hopping can be mapped to eight conjugations of ladders of CNOTs.}
    \phantomsection
    \label{fig:JW}
\end{figure}

We will use the simplest single hopping term to explain how to load USCC on the circuit by JW transformation and Trotterization.
\begin{align}
    &\exp\left[t_{ij} a_i^\dagger a_j - h.c.\right]\\
    &=\exp\left[t_{ij}(Q_i^-\prod\limits_{k=i+1}^L\sigma^z_k)(Q_j^+\prod\limits_{k=j+1}^L\sigma^z_k)-h.c.\right]\\
    &=\exp\left[t_{ij} Q_i^-(\prod\limits_{k=i+1}^{j-1}\sigma^z_k)(\sigma^z_j Q_j^+) - h.c.\right]\\
    &= \exp\left[it_{ij}/2\left[\sigma^x_i(\prod_{k=i+1}^{j-1}\sigma^z_k)\sigma^y_j - \sigma^y_i(\prod_{k=i+1}^{j-1}\sigma^z_k)\sigma^x_j\right]\right]\\
    &= H_i(HS^\dagger)^\dagger_j\exp\left[it_{ij}/2\prod\limits_{k=i}^j\sigma^z\right]H_i (HS^\dagger)_j \nonumber\\
    &\times(H S^\dagger)_i^\dagger H_j\exp\left[-it_{ij}/2\prod\limits_{k=i}^j\sigma^z\right](HS^\dagger)_i H_j
\end{align}
which is two conjugations of ladders of CNOTs shown in Fig. \ref{fig:JW}(a). Line 5 and 6 use $\exp(U^\dag MU) = U^\dag\exp(M)U$. The circuit of other hoppings under different condition can be derived using exactly the same method.

\begin{figure}[t!]
	\includegraphics[width=\linewidth]{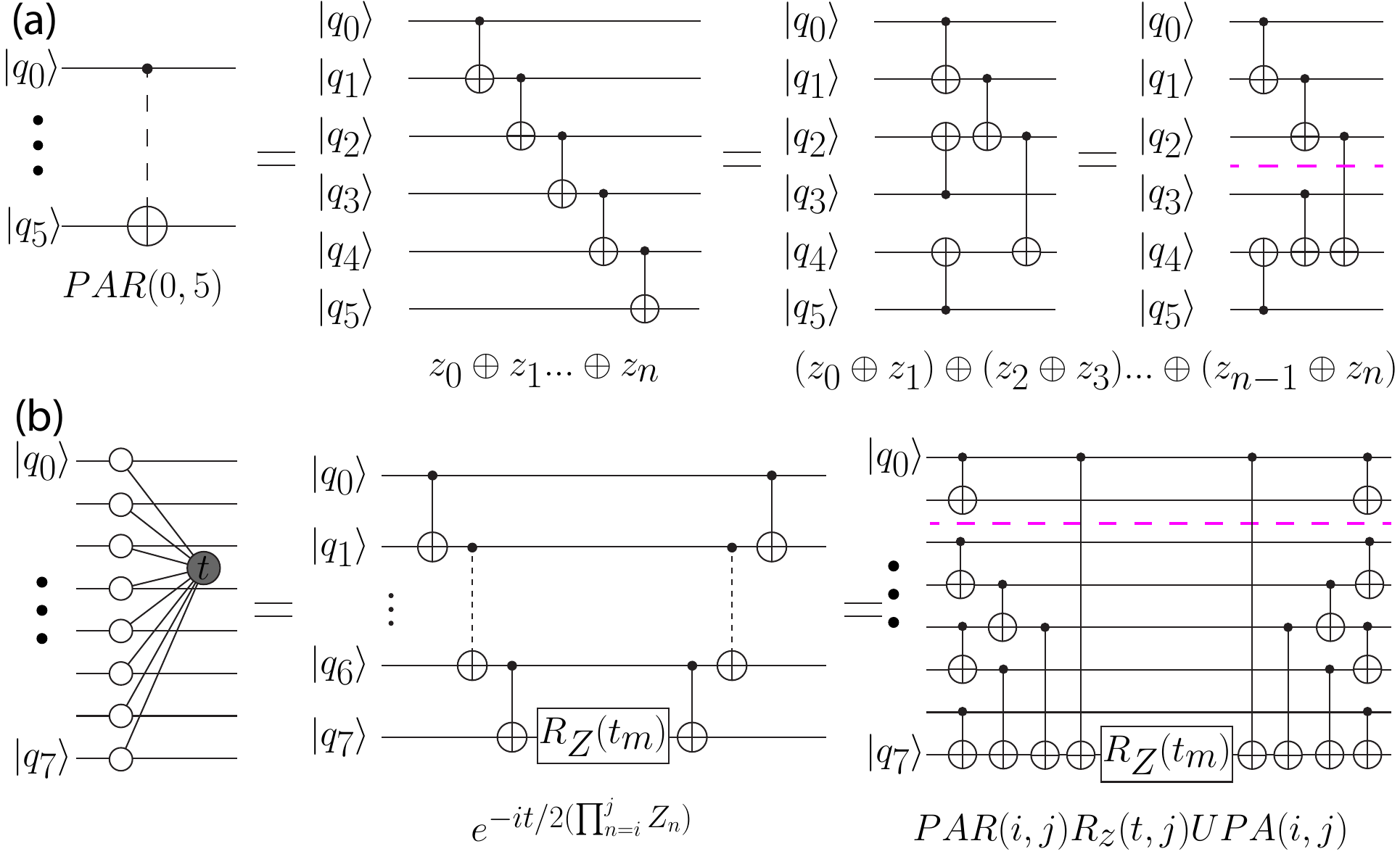}
	\caption{\red{
       \textbf{Parity check circuit}. (a) The parity check circuit (PAR) is represented as a dashed CNOT. This can be naively compiled as a CNOT cascade with the parity of all qubits on the check recorded at the last (target) qubit. The circuit depth can be reduced to $\log_2{(N_{checks})}$, and the circuit can be compiled with only one inter-module CNOT at each seam with additional depth $\leq N_{seams}$. (b) The ZX representation of a Pauli gadget $e^{-it/2\prod_{i=0}^7 Z_i}$, which can be trivially compiled to $R_z(t)$ conjugated by CNOT cascades, or $PAR(i,j)Rz(t,j)UPA(i,j)$. The seam is arbitrarily chosen between $\ket{q_1}$ and $\ket{q_2}$ for illustration.}
    }
    \phantomsection
    \label{fig:modular ladder}
\end{figure}

\begin{figure*}[t!]
	\includegraphics[width=0.8\linewidth]{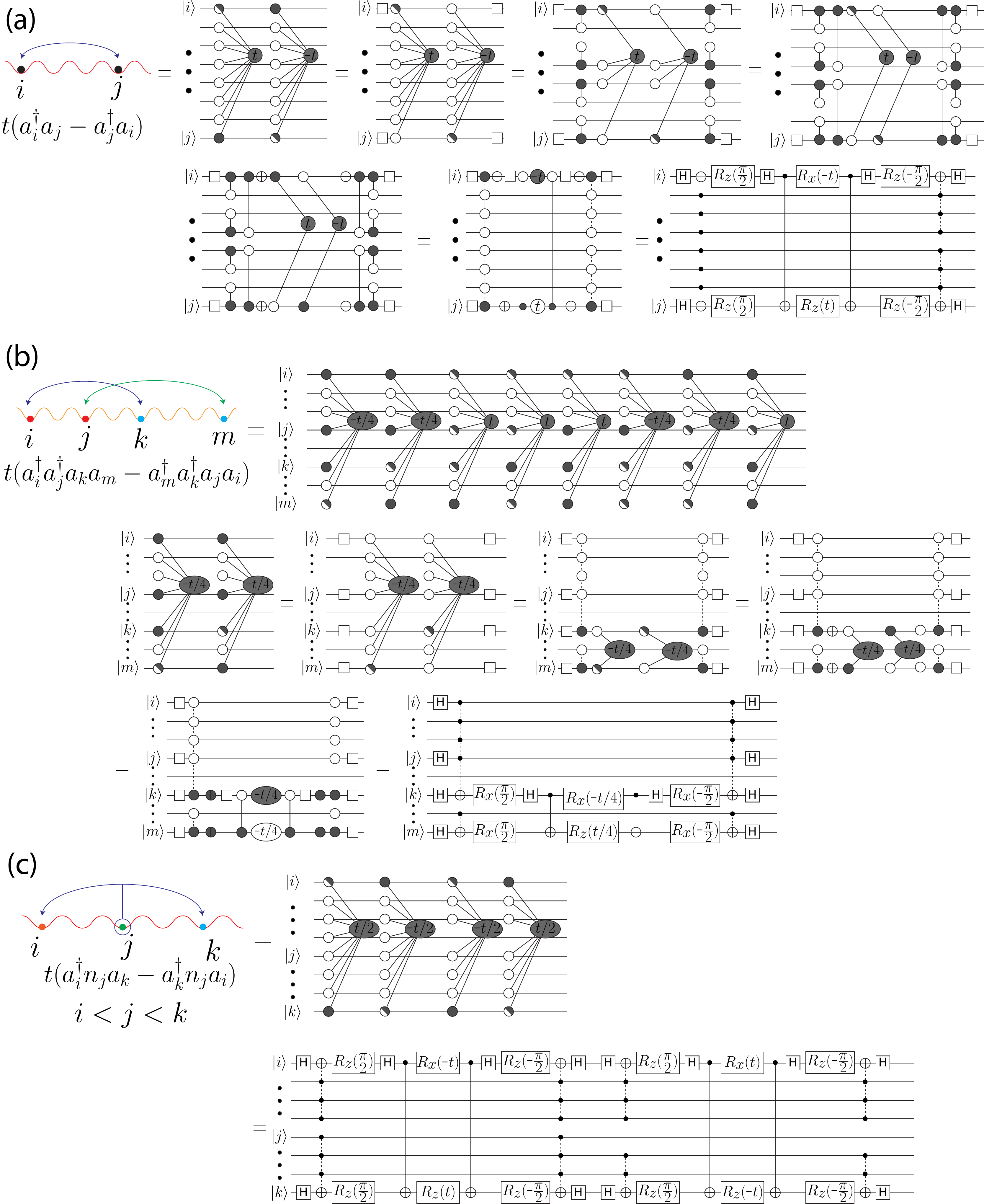}
    \centering
	\caption{\red{
        \textbf{ZX proof}. (a) The single hopping $e^{a_i^\dagger a_j - a_j^\dagger a_i}$ after the Jordan-Wigner transformation is two Pauli gadgets: $e^{it/2X_i(\prod_{k=i+1}^{j-1}Z_k)Y_j}$ and $e^{it/2Y_i(\prod_{k=i+1}^{j-1}Z_k)X_j}$. These two gadgets can be merged to two phase gates conjugated by parity check circuits. The ZX representation mostly follows \cite{Cowtan2020} with black dots representing X, white dots representing Z and mixed dots for Y. The last figure is the explicit circuit diagram. The black dots on the dashed line are the qubits checked by the PAR and the NOT gate presents the target qubit of the PAR. (b) Double hoppings can be represented by 8 Pauli gadgets. Each pair of gadgets with exactly 2 different basis can be merged. The second row is the example of merging $e^{X_i (\prod Z)X_j X_k(\prod Z) Y_m}$ and $e^{X_i (\prod Z)X_j Y_k(\prod Z) X_m}$. The last figure is the explicit circuit. (c)
        Controlled hopping can be represented by 4 Pauli gadgets. The last figure is the explicit circuit. In the middle of the circuit, the two columns of H gate conjugated with parity check circuits can be simplified to a single CNOT at the price of less commutativity. This technique also applies to double hoppings or to Pauli gadgets from different hoppings, which can significantly reduce the circuit depth but at the price of less commutativity. We leave the work of finding a good packing/merging algorithm to the future.
        }
    }
    \phantomsection
    \label{fig:zx proof}
\end{figure*}

\red{
\section{ZX hopping circuit}
\label{appendix: zx proof}
All dUSCC hoppings can be mapped to Pauli strings by the Jordan-Wigner transformation, and each Pauli string is naively compiled to the circuits shown in Appendix \ref{USCC loading}. The number of gates is halved using the ZX simplification~\cite{Cowtan2020}. 

In this paper, we use a very similar ZX proof in a more transparent way as shown in Fig. \ref{fig:zx proof}, which merges two Pauli ``gadgets'' into one circuit tile and halves the number of gates. The dashed line CNOT is the parity check circuit (PAR) shown in Fig. \ref{fig:modular ladder}(a). At the last (target) qubit, PAR records the occupancy of all sites on the Pauli string to track the fermionic statistics, and the reverse parity check circuit (UPA) at the end of each Pauli gadget unentangles all sites as shown in Fig. \ref{fig:modular ladder}(b). As phase gates only act on the target qubits, the order of checks can be freely changed, and the shortest circuit depth is $\log(N_{checks})$, where $N_{checks}$ is the number of qubits on the PAR. However, the optimal circuit can have at most $N_{checks}/2$ inter-module CNOTs at one seam. This can be reduced to one by reordering the order of checks as shown in Fig. \ref{fig:modular ladder}(b), where the seam is arbitrarily positioned between $\ket{q_1}$ and $\ket{q_2}$. In our compilation, we merge two Pauli gadgets into one circuit tile and effectively half the number of CNOTs while the number of inter-module CNOTs is also halved. The number of CNOTs and the circuit depth can be further reduced as suggested in the controlled hopping circuit in Fig. \ref{fig:zx proof} (c). The almost same PARs can be reduced to a single CNOT and half the circuit depth. However, merging more Pauli gadgets into a single tile will lead to less pseudo-commutativity, and we leave the optimal merging/packing algorithms to future work.
}

\section{Gate time benchmarking}
\label{benchmark}

\begin{figure*}[t!]
	\includegraphics[width=\linewidth]{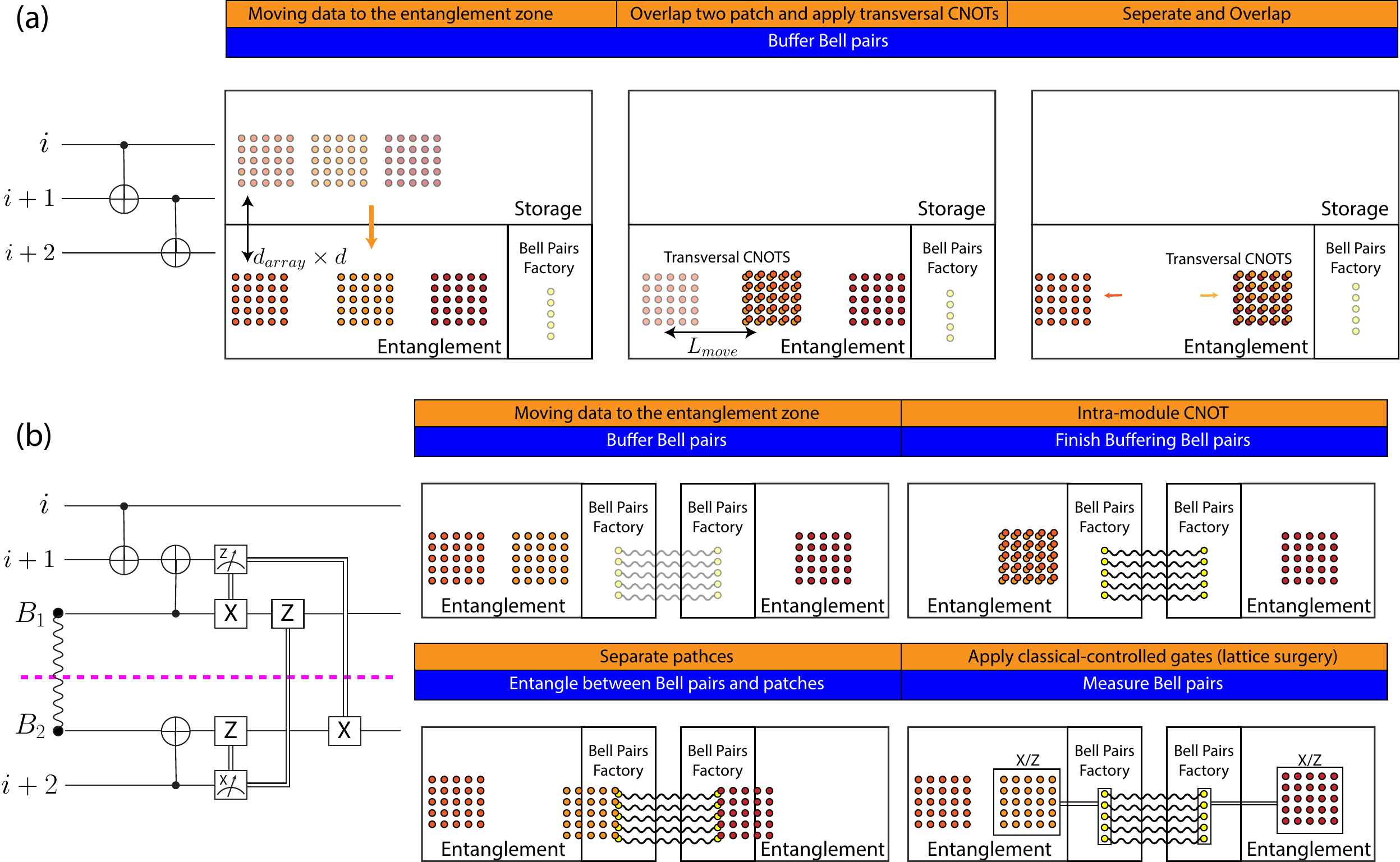}
	\caption{
        \textbf{Realization with reconfigurable atom arrays}. Each QPU is divided into three zones: qubit storage, entangle zone, and  Bell pairs factory
        (a) Intra-module ladder of CNOTs protocol. The orange bar represents the current intra-module operations, and the blue bar is the current inter-module operations. Qubits are moved from the storage zone to the entanglement zone by distance $d_{array}\times d$. Patch $i$ moves by distance $L_{move}$ and overlaps with patch $i+1$.  The logical CNOT is applied by a transversal CNOT between two patches. Patch $i$ moves back to the original position while patch $i+1$ overlaps with patch $i+2$ followed by a transversal CNOT. The Bell pair generation is in parallel with all these intra-module operations (b) Inter-module ladder of CNOTs protocol with lattice surgery. The intra-module part is exactly the same as (a). To execute the inter-module CNOT, patch is entangled with the Bell pairs. Classical controlled gates are applied on two patches based on the measurement results. In our assumption, we only apply 1 round of lattice surgery instead of d rounds.
    }
    \phantomsection
    \label{fig:hardware}
\end{figure*}

To give some context for the viability of performing modular algorithms, we focus specifically on the neutral atom array hardware platform. We consider logical qubits encoding with the rotated surface code with code distance $d = 5$. We assume that Bell pairs are generated between atoms in separate modules via photonic interconnects based on high-fidelity atom-photon Bell pairs and photonic Bell state analyzers~\cite{Covey2023,LiThompson2024,Li2025,VanLeent2022,Hartung2024}. Although the details of remote entanglement methods are outside the scope of this work, our assumed approach is expected to generate single remote Bell pairs at a rate $\gamma_\text{Bell}$, and thus the time required to obtain $N_\text{Bell}$ Bell pairs is $t_\text{Bell}(N)=N/\gamma_\text{Bell}$.

We assume that local one- and two-qubit logical gate operations within a module (``intra-module") are transversal, meaning that the constituent physical gate operations are performed in parallel for all physical qubits in the logical qubit. Hence, the intra-module bare gate time $t_\text{intra}^\text{bare}$ depends only weakly on $d$. We assume $t_\text{intra}^\text{bare}\approx0.5$ $\mu$s, which is typical in neutral atom processing hardware. We assume a zoned architecture in which distinct spatial regions are assigned for single-qubit operations, ancilla storage, two-qubit operations, and measurement~\cite{Bluvstein2022,Bluvstein2023}. Figure~\ref{fig:hardware} shows this vision.

The effective time required for an intra-module gate $t_\text{intra}$ will be dominated by the time required to reconfigure the array such that the logical qubit(s) are in the spatial zone associated with that operation. We assume that each module has $200\times200$ physical qubits and an array spacing of $d_\text{array}=5$ $\mu$m such that the full array length is $L=1$ mm in each direction. The entanglement zone can accommodate one row of data qubit patches (or 40 logical qubits). We assume that atoms are moved with an average speed of $v_\text{avg}=1$ m/s, limited by the tolerable inertial acceleration and deceleration associated with the motion profile. Hence, the time required to execute one ladder of CNOTs ($N_L^{\text{row}} = 40-1 = 39$ intra-module CNOTs) is

\begin{align}
t_\text{intra}^{\text{ladder}} &= \frac{d_\text{array}\cdot d+L_\text{move}\cdot N_L^{\text{row}}}{v_\text{avg}} + t^{\text{bare}}_{\text{intra}}\cdot{N_L^{row}}\\
&\approx \frac{L_\text{move}\cdot N_L^{\text{row}}}{v_\text{avg}}.
\end{align}
Here we calculate the time to evaluate longest intra-module ladder of CNOT that can be accommodated in the entanglement zone. In a more general case where the ladder of CNOT is shorter, more than 1 ladder of CNOTs can be applied simultaneously in the entanglement zone, and the total time is the same. 
The first term is the time required to move all atoms to the entangle zone. We assume there is no gap between the storage zone and the entanglement zone, and we move 40 $5\times 5$ (a $200 \times 5$ patch) patches simultaneously. The second term is the time to sequentially apply the nearest neighbor CNOT, and each patch moves $L_{\text{move}} = d\cdot d_{\text{array}}=5\cdot 5 \mu m$ to overlap with the neighbor patch. The third term is the time to do the transversal CNOTs between two logical patches. The time to apply one ladder of CNOTs is dominated by the time to sequentially overlap the neighbor patches. This time is $N_{L}^{\text{row}} = 39\times$ longer than the time to move the logical patches to the entanglement zone and $78\times$ longer than the time to apply transversal CNOTs. Therefore, all the operations that can happen in parallel (moving atoms to the entanglement zone and the transversal CNOTs) are negligible compared with the sequential overlapping between patches. Most assumptions of the parallelizable operations are included for simplicity and elucidation, and relaxing them, such as increasing the gap between the storage zone and the entanglement zone, only slightly changes the average time of intra-module CNOTs. The average time to apply one intra-module CNOT is
\begin{equation}
t_{\text{intra}}\approx\frac{t^{\text{ladder}}_{\text{intra}}}{N^{\text{row}}_L} = 25 \mu s.
\end{equation} 

\begin{figure}[t!]
	\includegraphics[width=\linewidth]{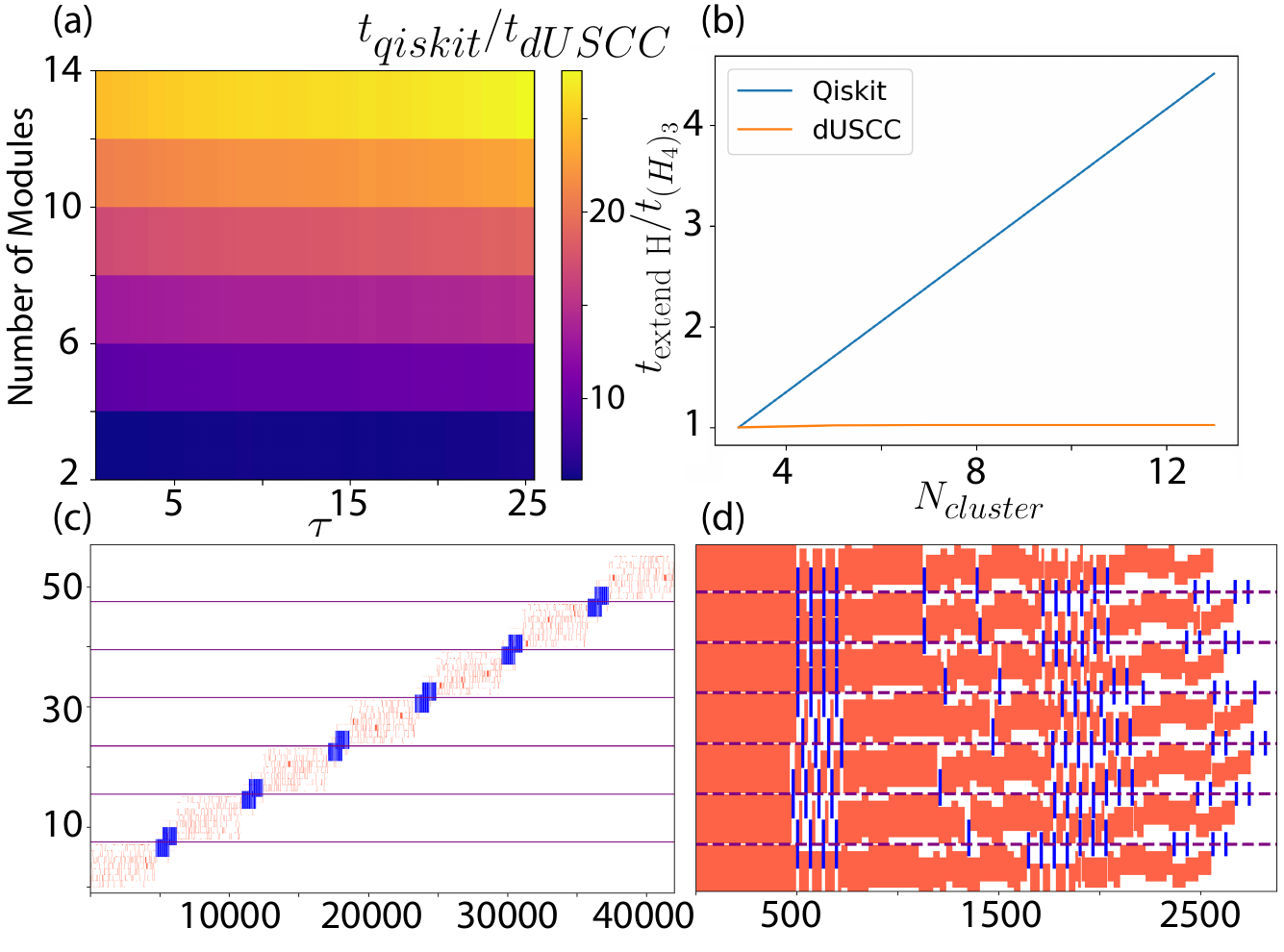}
	\caption{
       \textbf{The extended H$_2$ chain}. (a) The color plot of $t_{qiskit}/t_{dUSCC}$ at \red{$\epsilon =\times10^{-3}$}. \red{dUSCC is at most $\sim 35 \times$ shorter than Qiskit, which comes from a $2\sim3\times$ circuit depth saving from ZX compilation and $N_{cluster}$ saving by utilizing the geometry of the H-chain} (b) The time ratio of extended H-chain with different $N_{cluster}$ at \red{$\epsilon = 10^{-3}$}. The circuit time of the naive Qiskit compilation increases linearly with the number of modules whereas the circuit time of the dUSCC remains constant. (c),(d) are the abstract circuits of $(\text{H}_4)_7$ at $\tau = 10$. (c) The circuit of the naive compilation from Qiskit suggests Qiskit doesn't use the pseudo-commutativity to maximize the parallelism. (d) The circuit from dUSCC utilizes the translation symmetry of the H-chain, so the circuit time remains unchanged in the extended H-chain. Even without the translation symmetry, we believe the advantages of dUSCC is more dominant as the number of modules increases. 
    }
    \phantomsection
    \label{fig:extended H-chain}
\end{figure}

We assume two QPUs are photonically connected with Bell pair generation rate $\gamma_\text{Bell}=10^5/s$. The inter-module gates are applied by lattice surgery with each inter-module CNOT consuming $5$ Bell pairs. We assume the Bell pair storage is exactly enough to execute one inter-module tile (two inter-module CNOTs), and the buffering of Bell pairs can happen in parallel with all intra-module operations but not inter-module opereations. The time of inter-module CNOTs is the same as that of the intra-module CNOT if the Bell pairs are buffered.

Here, we use a very optimistic assumption that all Bell pairs and measurement are perfect and that the measurements time is negligible. With imperfect measurements, the lattice surgery becomes unstable and requires $d$ rounds of measurement. The exact cost of inter-module CNOTs should be calculated independently for each different platform and each set of assumptions.

\section{Extended H-chain system}
\label{extended H-chain}
The algorithm we adopt to generate all excitations in this paper has complexity $O(e^{N})$, but this is reduced to $O(poly(N))$ in the recent work~\cite{Verma2025}. To simulate the dUSCC in a larger molecular system, we stack the same $\text{H}_4$ modules in $(\text{H}_4)_3$ to extend the system in Fig. \ref{fig:schematic}(a). We simulate the extended H-chain system at $\epsilon = 10^{-3}$. \red{The dUSCC is at most $\sim35\times$ shorter than the Qiskit circuit as shown \ref{fig:extended H-chain}(a)}. The circuit compiled by Qiskit does not use pseudo-commutativity to optimize parallelism, so the circuit time increases linearly with the number of modules in Fig. \ref{fig:extended H-chain}(b). \red{The compact ZX compilation also reduces the circuit depth by $2\sim 3$ times compared with Qiskit}. dUSCC circuit also reflects the translation symmetry of the H-chain, and the circuit time is invariant in the extended H-chain \red{with different $N_{clusters}$}. Fig. \ref{fig:extended H-chain}(c) and (d) show the abstract circuit of $(\text{H}_4)_7$ from Qiskit and dUSCC.

\begin{figure}[t!]
	\includegraphics[width=\linewidth]{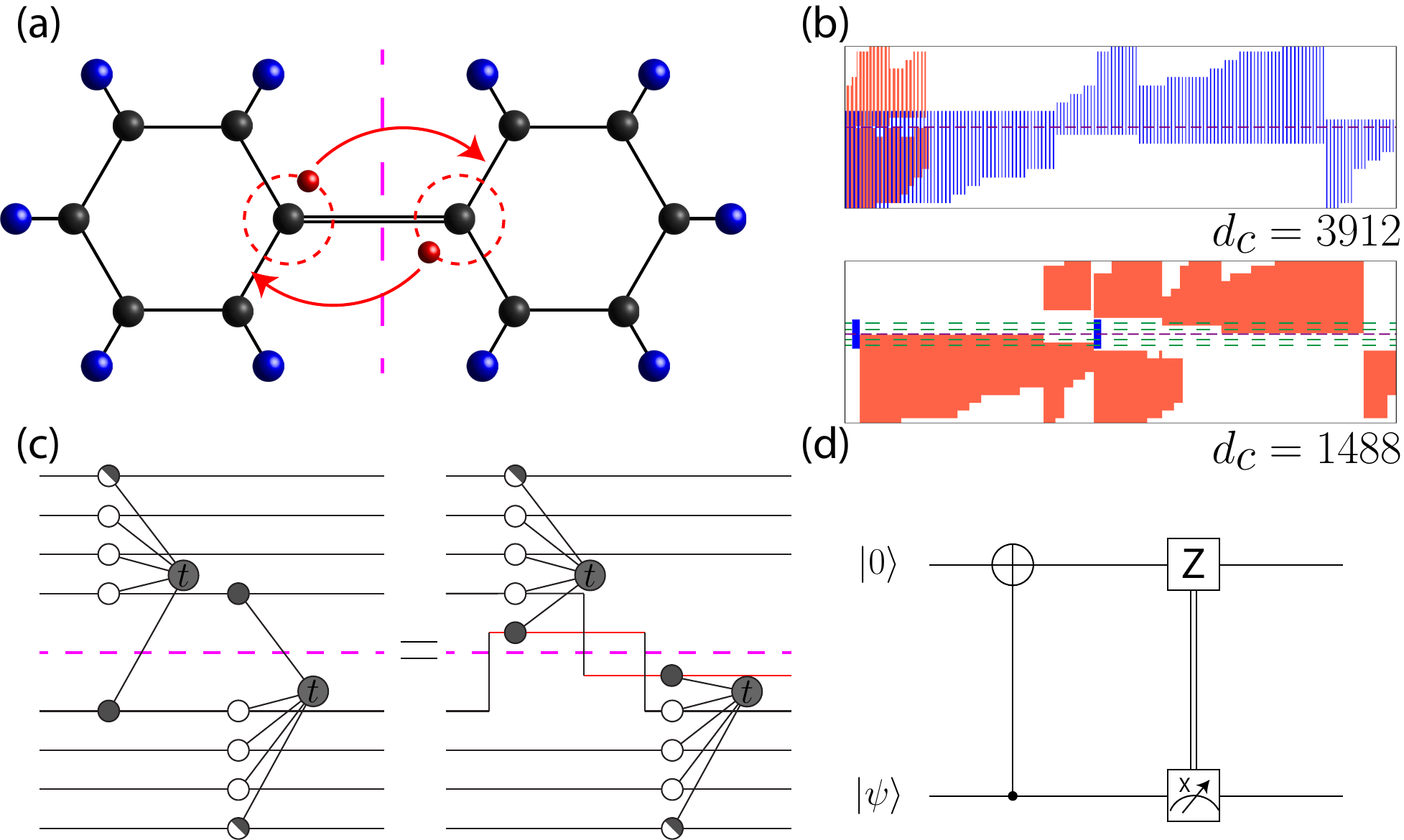}
	\caption{\red{
       \textbf{dUSCC on stilbene}.
    (a) The molecular structure of stilbene. The blue balls are hydrogen atoms and black balls are carbon atoms. The dashed line represents the dUSCC anstaz seam. Most inter-module communications are from the C-C double bond electrons hopping from one benzene ring to another. (b) dUSCC pakcing result for the stilbene at $\epsilon = 5\times10^{-3}$ and $\tau = 10$. The top figure is the dUSCC pakcing result using the gate-teleportation where most tiles are inter-modular. The bottom figure is the dUSCC packing using qubit-teleportation where the green dashed lines are the positions of ancilla qubits in $\ket{0}$ state. Each blue tile consists of two qubit-teleportation (two inter-module CNOTs). The dUSCC of the stilbene only requires four inter-module CNOTs, so the latency of inter-module communication is ignorable. (c) The ZX representation of the qubit-teleportation of Pauli gadgets. As most inter-module communications origin at a single electron, the inter-module communications can be significantly reduced using qubit-teleportation. The red lines are the additional ancilla qubits required. (d) The explicit circuit of qubit teleportation.
    }}
    \phantomsection
    \label{fig:stilbene}
\end{figure}

\begin{figure}[t!]
	\includegraphics[width=\linewidth]{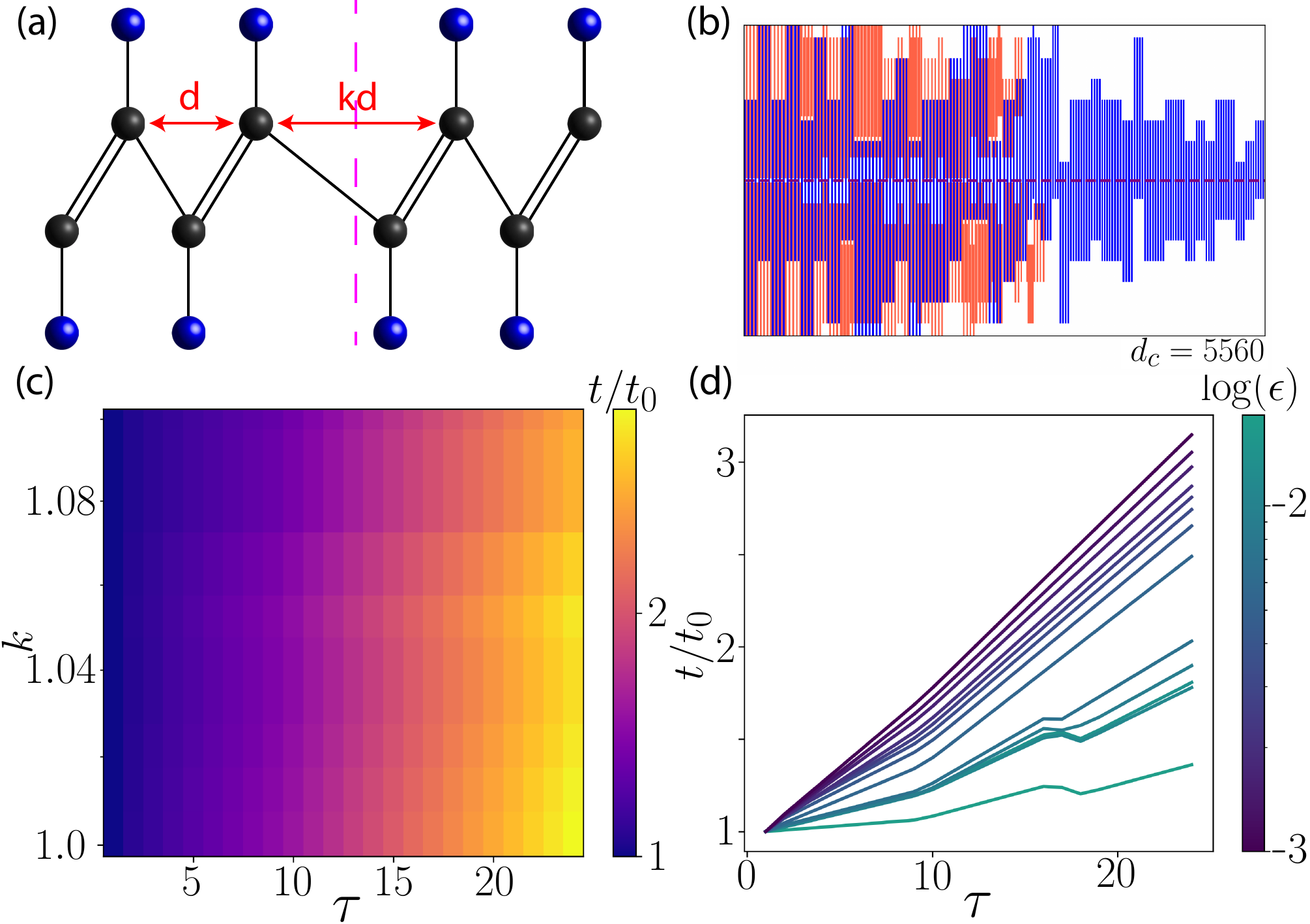}
	\caption{\red{
       \textbf{dUSCC on polyene}. 
       (a) The molecular structure of polyene, which has the alternative C-C single bonds and double bonds.  The blue balls are hydrogen atoms and black balls are carbon atoms. The dashed line represents the seam of dUSCC ansatz. The middle single bond is adjusted to reduce the inter-module entanglement. (b) The dUSCC packing result of polyene at $\epsilon = 5\times10^{-3}$ and $k = 1.1$. (c) The polyene packing result at $\epsilon = 10^{-3}$. The circuit time of dUSCC increases linearly with the inter-module latency, and the delay of the circuit also depends on the inter-module entanglement. The dUSCC is less delayed as the distance between two clusters increases (d) The circuit time at $k = 1.1$. The inter-module latency becomes more significant as $\epsilon$ decreases and inter-module entanglement becomes stronger.
    }}
    \phantomsection
    \label{fig:polynene}
\end{figure}

\red{
    \section{dUSCC on other molecules}
    \label{appendix:more molecules}
    Beyond the H-chain, we also apply dUSCC on the  stilbene and polyene molecules, (Fig. \ref{fig:stilbene} and Fig. \ref{fig:polynene}), which have been previously used as benchmarks~\cite{doi:10.1021/acs.jpca.8b01554} and where USCC can approximate the ground state energy within the chemical accuracy~\cite{Shreya2025}. Stilbene has a structure of two benzene rings connected by a C-C bond, and most inter-module connections are from the interactions between one electron on the C-C bond and another on the benzene ring (Fig. \ref{fig:stilbene}(a)). Although extensive inter-module communications are required when gate-teleportation is used (see top figure of Fig. \ref{fig:stilbene}(b)), the communication is ignorable (four inter-module CNOTs) using qubit-teleportation at the cost of 4 more ancilla qubits (bottom figure of Fig. \ref{fig:stilbene}(b)). The qubit teleportation in both ZX representation and explicit circuit is shown in Fig. \ref{fig:stilbene}(c,d).
    
    We also apply dUSCC to another chain-shaped molecule, polyene, which has alternative C-C double bonds and C-C single bonds  [Fig. \ref{fig:polynene}(a, b)]. We stretch the middle bond ($d_{inter} = kd$) to vary the inter-module interactions, and the latency of dUSCC decreases as the distance increases as shown in Fig. \ref{fig:polynene}(c). dUSCC is delayed more as $\epsilon$ decreases as shown in Fig. \ref{fig:polynene}(d), which agrees with our observations in the H-chain molecules and further verifies that dUSCC behaves better in a system with sparse inter-module connections but with complicated intra-module connections. 
    
    In this paper, we focus on chain-structure molecules and stilbene, but there are more molecules with sparse inter-module connections compared with intra-module interactions. Stilbene also demonstrates a situation where the qubit-teleportation significantly outperforms the gate-teleportation in dUSCC. We expect other molecules with similar structures to have the same properties, and there are more undiscovered molecular structures suitable for modular algorithms. Our research motivates more research on the molecules that would benefit from fragmentation and modularization and we leave this to future work.
}

\begin{figure}[t!]
	\includegraphics[width=\linewidth]{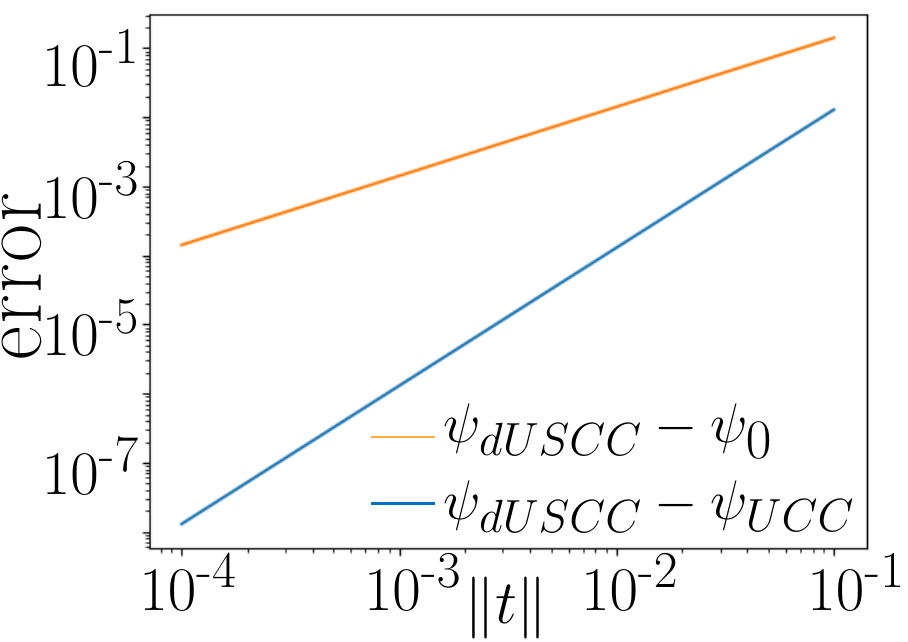}
	\caption{\red{
       \textbf{Circuit equivalency between the dUSCC and the Qiskit} The comparison between dUSCC and qiskit USCC on the stilbene with 20 active spaces at $\epsilon = 5\times10^{-3}$. To simulate the most general case, dUSCC and qiskit USCC circuit use the same and random parameters. Two circuits are separately applied on the same random initial state, and the state difference between the dUSCC and qiskit USCC is in in the order of $O(\|t\|^2)$, the same as Trotter error. The reference line (orange line) shows the difference between dUSCC result state and the initial state is $O(\|t\|)$, so our observation of $\|t\|^2$ error is from the Trotter error.
    }}
    \phantomsection
    \label{fig:Trotterization}
\end{figure}

\red{
    \section{Pseudo-commutativity and circuit equivalency}
    \label{appendix: pseudo-commutativity}
    By Trotterization, the sum of Hamiltonian terms on the exponents can be approximated by the products of exponentiation of each Hamiltonian terms. In this paper, we only consider the first-order Trotterization,
    \begin{equation}
        e^{i\mathcal{H}t} = e^{it\sum_{k}h_k} = \prod_k e^{ith_k} + O(t^2\max_{i,j}([h_i, h_j]).
        \label{Trotterization}
    \end{equation}
    As the error includes the commutation error and the scaling of Trotter error is independent of the order of product of each $e^{ih_k}$, all terms can be reordered without changing the scaling of the error, and we define this as pseudo-commutativity.

    To explicitly show this pseudo-commutativity, we run both dUSCC and Qiskit USCC on the stilbene at $\epsilon = 5\times10^{-3}$. We first collect all the selected hoppings from USCC and feed the hoppings to both Qiskit and dUSCC. To prove this pseudo-commutativity in the most general case, we evolve two circuits with the same but random parameters on a random initial state, and Fig. \ref{fig:Trotterization} shows a $\|t\|^2$ Trotter error. Therefore, in the context of Trotterization, Qiskit circuit and the dUSCC circuit are ``equivalent'' as they have the same scaling error. The stilbene system simulation requires 20 qubits, so finding the exact error and comparing the overhead of Trotter error numerically is outside the scope of this work.

    The order of terms can affect the overhead of the Trotter error. However, both Qiskit and dUSCC do not consider to reduce the overhead of the Trotter error by changing the order of hoppings. Putting commutative terms together and introducing each $e^{ih_k}$ term probabilistically can significantly reduce the overhead \cite{Ouyang2020, Cambell2019, Childs2019}. Our technique already combines more than half of the commutative terms by merging two commute Pauli strings into a single tile as shown in Appendix \ref{appendix: zx proof}, and the probabilistic introduction of each $e^{ih_kt}$ term works very similarly to the dUSCC selection, so dUSCC should also fit the more advanced Trotterized algorithm, but more research is needed to find the balance between error overhead and parallelism.
}

\section{FeMoco Number of Modules}
\label{app:FeMoco}
In the main text, we demonstrate the efficacy of the dUSCC algorithm on model Hamiltonians. To provide context for what is
required for utility-scale problems in quantum chemistry, here we briefly review the estimated resources required to
compute the iron molybdenum cofactor of nitrogenase, commonly known as FeMoco~\cite{reiher2017elucidating,li2019electronic}. This problem is of practical interest for designing
catalysts for nitrogen fixation and has been used as a standard benchmark of progress in the 
quantum algorithms. The Hamiltonian of Ref.~\cite{reiher2017elucidating} consists of 54 orbitals
and, using Trotterization within quantum phase estimation, requires 111 logical qubits and around 
$10^{15}$ logical T-gates to find the ground state energy~\cite{reiher2017elucidating,otten2023qrechem}.
Advances in quantum algorithms, including qubitization~\cite{low2019hamiltonian}, tensor hypercontraction~\cite{lee2021even}, and spectral 
amplification~\cite{low2025fast} have reduced the resource costs, resulting in estimates of 
1137 logical qubits executing $3.4\times10^8$ logical T-gates. Such gate depths, whether using Trotterization
or qubitization, necessitate high distance surface codes, which will depend sensitively on the error model
and physical architecture~\cite{mohseni2024build}.

To provide a first-pass estimate of the number of modules required for a utility-scale application,
we will assume that 111 logical qubits encoded at surface code distance $d_s=50$ are required.
We do not provide explicit circuits or T-gate counts here; our goal is to get an approximate
number of modules. Since a
surface code requires approximately $2d_s^2$ qubits, this gives a physical qubit count of $O(500k)$ 
physical qubits. Additional physical qubits would be needed for T-gate factories, which themselves
can take a large amount of physical qubit count~\cite{mohseni2024build}. 
A conservative estimate of the number of atoms that could be stored in one 
module is $O(10k)$~\cite{Manetsch2024}. Thus, $\gtrsim$50 modules would be required to store and process these logical qubits. If we alternatively assume that an atomic processor can accommodate $O(50k)$ physical qubits, then $\gtrsim$10 modules are still required.
\begin{figure}[t]
    \includegraphics[width=\linewidth]{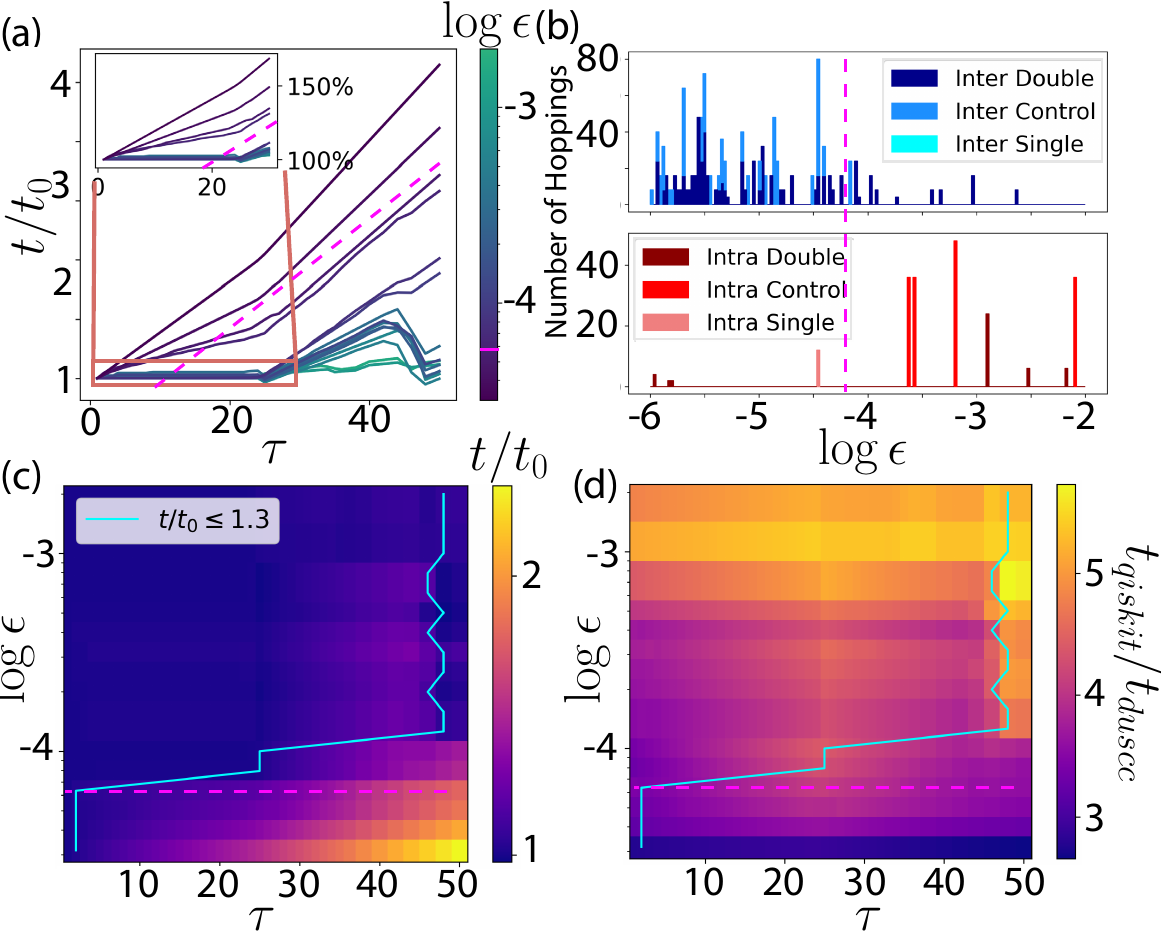}
	\caption{
        \textbf{Results}. \red{All results are from the dUSCC simulation in the $(H_4)_3$ at $d = 3d_0$ and logical single qubit rotation is $40\times$ as expensive as logical CNOTs. (a) dUSCC circuit time compared to the no-latency case, $t/t_0$, with different inter-module gate costs $\tau$ and different selection criteria $\epsilon$.
        (b) The hopping diagram of $(\text{H}_4)_3$ over the selection criteria. (c) The phase diagram of the delayed dUSCC time. The blue line captures the transition from the ``free'' dUSCC with $t/t_0 < 1.3$ as the largest fluctuation in (a) is $t/t_0 = 1.3$. The existence of the ``free'' modularization is also captured by the pink line $\epsilon = 10^{-4.2}$. dUSCC of $(\text{H}_4)_3$ is free at $\epsilon \leq 10^{-4}$ at $\tau \leq 48$, and the transition between the existence of ``free'' modularization is sharper.  (d) The time ratio between the dUSCC compilation and the Qiskit compilation at. Our compilation also shows its best advantages at the blue boundary in (c).}
    }
    \phantomsection
    \label{fig:dw40}
\end{figure}

\section{Double Packing Scheme}
\label{Appendix: Double Packing}

In this section, we will introduce the double packing algorithm in more details. 

\red{
    \textbf{Assumptions}. In this paper, we use very optimistic intra-module assumptions (or very pessimistic inter-module assumptions) and ignore the cost of all Clifford gates. Although logical Pauli gates can be applied classically at no cost by changing the Pauli frame \cite{Chamberland2018}, the Clifford gates in the surface code such as $H$ and $S$ require a rearrangement of qubits inside the code patch \cite{Chen2024}, which takes half the time of a logical CNOT. A logical rotation gate with arbitrary error $\epsilon$ can be decomposed to roughly $10 \log{\frac{1}{\epsilon}}$ T gates \cite{Ross2016}, and even for the simplest $(H_4)_3$ system, the gate error $\epsilon \leq 10^{-4}$ is necessary, which translates to at least 40 T gates per phase rotation gate with 1 T gates taking roughly the same time as $1$ or $0.1$ (with fast non-local CCZ gates) logical CNOT using code cultivation \cite{Gidney2024, Chen2024RP2, Sahay2025}. Given the fast development of the code cultivation, we optimistically assume the logical phase rotation is $10\times$ more expensive than a CNOT. A stricter assumption like phase rotation is $40\times$ more expensive than logical CNOTs can significantly improve our results as shown in Fig \ref{fig:dw40}. The threshold of ``free'' modularization is raised to $\tau = 48$ at $\epsilon \leq 10^{-4}$. The boundary of the existence of ``free'' modularization of dUSCC is sharper when logical phase rotations are $40\times$ as expensive as the logical CNOTs.
}

\begin{algorithm}[H]
\caption{Double Packing}
    \begin{algorithmic}[1]
    \Require Tile list \texttt{Tilelst} as \texttt{[width, height, x, y]}, seam positions, and inter-module gate cost $\tau$
    \Output PlacedTileList as \texttt{[placedx, width, height, x, y]}
    \State Sort \texttt{Tilelst} by descending height
    \State Initialize \texttt{InterGrid} and \texttt{IntraGrid}
    \For{each \texttt{Tile} in \texttt{Tilelst}}
        \If{\texttt{Tile} is intra-module}
            \For{each $x < x_{\max}$}
                \If{\texttt{Tile} fits in \texttt{IntraGrid} at $x$}
                    \State Mark \texttt{IntraGrid} as occupied by \texttt{Tile}
                    \State Mark \texttt{InterGrid} as occupied by \texttt{Tile}
                    \State Append \texttt{Tile} to \texttt{PlacedTileList}
                    \State \textbf{break}
                \EndIf
            \EndFor
        \ElsIf{\texttt{Tile} is inter-module}
            \State Adjust $\tau$ so that following intra-module tiles fit
            \State maskedTile $\gets [\text{width} + 2 \times (\text{height} - 1) \times (\tau - 1), \text{height}, x, y]$
            \For{each $x < x_{\max}$}
                \If{\texttt{maskedTile} fits in \texttt{InterGrid} at $x$}
                    \State Mark \texttt{InterGrid} as occupied by \texttt{maskedTile}
                    \State Mark \texttt{IntraGrid} as occupied by \texttt{Tile}
                    \State Append \texttt{Tile} to \texttt{PlacedTileList}
                    \State \textbf{break}
                \EndIf
            \EndFor
        \EndIf
    \EndFor
    \State \Return \texttt{PlacedTileList}
    \end{algorithmic}
\end{algorithm}

\textbf{Circuit Tile Compilation}. By leveraging the pseudo-commutativity of the Trotterization, maximizing the parallelism in dUSCC can be solved by a 1+1D classical packing problem. All non-commutative smallest components are compiled to tiles using the following rules:
\begin{enumerate}
    \item The circuit tile $M_{ij}^\dagger\exp[t_{ij}\prod_k\sigma_k^z]M_{ij}$ spans from i to j with the circuit depth as its width.
    \item Due to the pseudo-commutativity of USCC ansatz, each circuit tile can move horizontally.
    \item To maintain the O($N_{cluster}$) overhead of JW transformation, no index reordering is allowed, so tiles cannot move vertically.
    \item The circuit depth can be treated as circuit time with one circuit depth representing time to execute one intra-module CNOT. Each inter-module tile includes exactly $2(k-1)$ inter-module gates (k is the number of modules the tile span). The delay of inter-module CNOT is effectively separating inter-module circuit tiles by $2(k-1)(\tau-1)$.
\end{enumerate}

\textbf{Double Packing Scheme}. Here, we present a heuristic packing algorithm using the greedy algorithm. On the high level, double packing algorithm packs the two grids, InterGrid and IntraGrid, simultaneously where only the InterGrid can see the inter-module latency. If an intra-module tile fits in the IntraGrid, it occupies both grids. If an inter-module tile masked with separation fits in the InterGrid, the masked tile occupies the InterGrid, and the original tile occupies the IntraGrid. The separation $\tau$ is adjusted so that there always exists an intra-module tile fit between inter-module tiles to maximize the parallelism. The pseudo-code is shown in the Algorithm below, and the C++ code of double packing is in the supplementary material. All code and raw data can be found on GitHub \cite{TianRepo}.

\clearpage

\end{document}